# Microstructural Design via Spinodal-Mediated Phase Transformation Pathways in HEAs using Phase-Field Modelling

Kamalnath Kadirvel[1a], Hamish L. Fraser[a], Yunzhi Wang[2a]

[a] The Ohio State University, Columbus, Ohio

**Abstract:** Even though the majority of the so-called high-entropy alloys (HEAs) are multi-phase rather than single-phase solid solutions at thermodynamic equilibrium, recent studies on HEAs do open up a massive compositional space for the exploration of novel microstructures with the potential of exhibiting greatly enhanced functional or mechanical properties. Understanding the phase transformation pathways (PTPs) and microstructural evolution in multi-phase HEAs will aid alloy and process designs to tailor the microstructures for specific engineering applications. In this work, we study microstructural evolution in two-phase HEAs where a disordered parent phase separates into a mixture of two phases: an ordered phase ($\beta'$) + a disordered phase ($\beta$) upon cooling following two different PTPs: (i) congruent ordering followed by spinodal decomposition in the ordered phase and then disordering of one of the ordered phases, i.e., $\beta \rightarrow \beta' \rightarrow \beta_1' + \beta_2' \rightarrow \beta + \beta_2'$ and (ii) spinodal decomposition in the disordered phase followed by ordering of one of the disordered phases, i.e., $\beta \rightarrow \beta_1 + \beta_2 \rightarrow \beta_1 + \beta'$. We systematically investigate the effects of equilibrium volume fractions of individual phases, free energy landscapes (in particular, the location of the critical point of the miscibility gap relative to the compositions of the final two equilibrium phases), and elastic modulus mismatch between the two equilibrium phases on the microstructural evolution of these HEAs. We focus on the following morphological characteristics: bi-continuous microstructures vs. precipitates + matrix microstructures, ordered matrix + disordered precipitates vs. disordered matrix + ordered precipitates, and the discreteness of the precipitate phase. This parametric study may aid in multi-phase HEA design for desired microstructures.

## 1. Introduction

High entropy alloys (HEAs) or multi-principal-element alloys (MPEAs) have gained a lot of attentions in the past two decades [1-10]. Although the stability of multicomponent solid solution was originally attributed to the high configurational entropy [6, 11], both enthalpy and entropy play a significant role in stabilizing a solid solution [12] and many of the single-phase HEAs are only metastable. While many of the original efforts in HEA development were focused on designing single-phase solid solutions, recent works are directed towards multi-phase HEAs [13-23] to develop new alloys for high temperature structural applications in extreme enviroments and also for enhanced magnetic properties. The efforts have been devoted to utilizing the metastability of the vast majority of HEAs to design novel two-phase or multi-phase microstructures for better functional or mechanical properties [24-27].

[1] Graduate Research Associate, Email: kadirvel.1@buckeyemail.osu.edu
[2] Email: wang.363@osu.edu

Metastable HEAs are likely to have a miscibility gap as they have large positive enthalpies of mixing [11, 28-30]. A recent study has showed the existence of spinodal regions in many HEA systems using CALPHAD modelling [31]. The two-phase (B2/bcc) microstructures observed in Al$_{0.5}$NbTa$_{0.8}$Ti$_{1.5}$V$_{0.2}$Zr [32], AlMo$_{0.5}$NbTa$_{0.5}$TiZr [16, 22] and AlCo$_{0.4}$Cr$_{0.6}$FeNi [27] are likely to have evolved through spinodal-mediated phase transformation pathways (PTPs) as suggested by the experiments. In AlMo$_{0.5}$NbTa$_{0.5}$TiZr [16, 22, 33] at the solutionizing temperature, an homogeneous bcc phase was observed. Upon furnace cooling, the alloy decomposes into B2 and bcc phases together with some grain boundary precipitates. Interestingly, the ordered phase (B2) was the matrix and the disordered phase (bcc) became discrete precipitates, unlike Ni-based superalloys where ordered precipitates are embedded in a disordered matrix. Similarly in Al$_{0.5}$NbTa$_{0.8}$Ti$_{1.5}$V$_{0.2}$Zr [32], at the solutionizing temperature, a homogeneous single phase (bcc or B2) was observed. Upon furnace cooling, a two-phase mixture of bcc and B2 was observed, again with bcc precipitates embedded in a B2 matrix. Soffa and Laughlin [34] have discussed several conventional alloys (i.e. with one or two principal elements) in both fcc-based and bcc-based crystal systems where both spinodal decomposition and order↔disorder transtitions occur in a cooperative or interdependent manner during the PTPs. They also show the schematic free energy curves of both the ordered and disordered phases which can lead to such PTPs. For instance, in an Al-Fe alloy, there can be a conditional spinodal where the spinodal decomposition is preceded by congruent ordering [34, 35]. Later Soffa and Laughlin explored the free energy curves where first-order order↔disorder transtition occurs in conjunction with the spinodal decomposition [36, 37]. The microstructural evolutions during precipitation of an ordered intermetallic phase from a disordered matrix phase through both first-order and second-order transitions have also been simulated by using microscopic diffusion theory [38]. Possible spinodal mediated PTPs in $\alpha/\beta$ two-phase Ti alloys was also discussed recently using the schematic free energy curves [39]. Similar or more complicated PTPs are expected for HEAs.

If the moduli of the decomposed phases were to be similar, spinodal decomposition would lead to bi-continuous or interconnected two-phase microstructures when the volume fraction of the two phases are close to each other [40]. In many HEAs, the ordered phase ($\beta'$) becomes the matrix and the disordered phase ($\beta$) becomes the precipitates [16, 27, 32, 41]. This is contrary to the typical microstructures observed in Ni-based superalloys where the ordered phase ($\gamma'$) becomes the precipitates and the disordered phase ($\gamma$) becomes the matrix. Typically, we expect the high temperature disordered phase to be the matrix where ordered precipitates occur by nucleation and growth. The ordered matrix phase and the highly discrete disordered precipitates observed in HEAs are explained based on the modulus mismatch [22, 32, 41, 42]. The high moduli phase prefers to be embedded within the low moduli phase during spinodal decomposition [43-45]. Based on this explanation, the modulus of the ordered phase must be significantly smaller than that of the disordered phase. However, for most refractory HEAs [22, 32, 41], it was neither shown through experiments nor first-principles calculations that the ordered phase indeed has a much lower modulus. It is possible that other factors can lead to the formation of highly discrete microstructures apart from the modulus mismatch (or in addition to the modulus mismatch). Even if the modulus of the ordered phase is higher, the magnitude of the modulus mismatch required to cause phase inversion is quite large ($C_{ijkl}^{ord}/C_{ijkl}^{dis} \approx 0.5 - 0.6$) [44] and is difficult to comprehend without proper experimental measurements. We demonstrate through phase-field simulations that the effect of the shape of the free energy curve can also contribute to the formation of an ordered matrix and highly discrete disordered precipitates, in addition to the modulus mismatch.

In this study, we model microstructural evolution in HEAs through spinodal-mediated PTPs using the phase-field method. We consider a homogeneous single-phase ($\beta$) disordered solid solutions at high temperatures, which decomposes into a two phase $\beta' + \beta$ microstructure. We consider two different PTPs for the decomposition: (i) congruent ordering followed by spinodal decomposition in the ordered phase and then disordering of one of the ordered phases, i.e., $\beta \rightarrow \beta' \rightarrow \beta'_1 + \beta'_2 \rightarrow \beta + \beta'_2$ and (ii) spinodal decomposition in the disordered phase followed by ordering of one of the disordered phases, i.e., $\beta \rightarrow \beta_1 + \beta_2 \rightarrow \beta_1 + \beta'$. We investigate systematically the effects of equilibrium volume fractions of individual phases, free energy landscapes (in particular, the location of the critical point of the miscibility gap relative to the compositions of the final two equilibrium phases), and elastic modulus mismatch between the two equilibrium phases on the microstructural evolution of these HEAs. In particular, we highlight the importance of the shape of the free energy curve in the microstructural evolutions in addition to modulus mismatch, which, to the authors' knowledge, has not been explored in the current literature.

## 2. Phase-field model

We consider two possible spinodal-mediated PTPs in the current work: (1) $\beta \rightarrow \beta' \rightarrow \beta'_1 + \beta'_2 \rightarrow \beta + \beta'_2$: (i) Congruent ordering of $\beta$ into $\beta'$; (ii) spinodal decomposition of ordered phase $\beta'$ into solute-lean $\beta'_1$ and solute-rich $\beta'_2$ phases; (iii) disordering of solute-lean $\beta'_1$ ordered phase into $\beta$. (2) $\beta \rightarrow \beta_1 + \beta_2 \rightarrow \beta_1 + \beta'$ : (i) Spinodal decomposition of $\beta$ into solute-lean $\beta_1$ and solute-rich $\beta_2$ phases; (ii) Chemical ordering of solute-rich $\beta_2$ into $\beta'_2$. Note that we chose the solute atoms to be $\beta'$ forming elements similar to our earlier work [45]. In general, the order $\leftrightarrow$ disorder transition can be first-order or second-order and based on this transition the topology of the phase diagram can be quite different as illustrated in Figure 1. Currently, we assume that the transition is first-order and in the subsequent work second-order transition will be considered.

To simulate the microstructural evolution along both the PTPs described earlier, we employ the phase-field model with two order parameters: (i) concentration $c(\vec{r}, t)$ that describes local variation in solute concentration and (ii) structural order parameter $\eta(\vec{r}, t)$ that differentiates between the $\beta$ phase ($\eta = 0$) and $\beta'$ phase ($\eta = 1$). The free energy of the system $F[c(\vec{r}, t), \eta(\vec{r}, t)]$ is formulated as a functional of order parameter field as follows

$$F[c(\vec{r}, t), \eta(\vec{r}, t)] = \int \left[ f(c, \eta) + \frac{\kappa_c}{2} (\nabla c)^2 + \frac{\kappa_\eta}{2} (\nabla \eta)^2 \right] dV + E^{el}[c, \eta] \quad (1)$$

where $f(c, \eta)$ is the local free energy density, $\kappa_c$ is the gradient energy coefficient of concentration, $\kappa_\eta$ is the gradient energy coefficient of the structural order parameter and $E^{el}[c, \eta]$ is elastic energy functional.

## 2.1. Chemical free energy

The local chemical free energy density $f(c, \eta)$ of the system can be constructed using the free energy of the disordered phase, $f^\beta(c)$, and the free energy of the ordered phase, $f^{\beta'}(c)$, as follows

$$f(c, \eta) = f^\beta(c)\big(1 - h(\eta)\big) + f^{\beta'}(c)h(\eta) + \omega g(\eta) \tag{2}$$

where $h(\eta) = \eta^2(3 - 2\eta)$ is an interpolation function, $g(\eta) = \eta^2(1 - \eta)^2$ is the double-well potential and $\omega$ is the barrier height. The order parameters ($c(\vec{r}, t)$ and $\eta(\vec{r}, t)$) and the free energy are non-dimensional in this work. All the free energy surfaces $f(c, \eta)$ used in the present work are shown in Figure 2 and the free energies $f^\beta(c)$ and $f^{\beta'}(c)$ that were used to construct $f(c, \eta)$ are shown in Figure 3. The non-dimensional solute concentration ($c$) can be related to real solute concentration ($x$) using $c = (x - x_\beta)/(x_{\beta'} - x_\beta)$ where $x_\beta$ and $x_{\beta'}$ are the equilibrium solute concentrations of the $\beta$ and $\beta'$ phases, respectively. Note that the non-dimensional solute concentration can be negative based on this definition. Throughout this work, the volume fraction $f_V$ refers to the equilibrium volume fraction of $\beta'$ unless indicated otherwise. We will simulate the microstructural evolution for the alloys only in the high-volume fraction regime, i.e., for the volume fractions being 40%, 50% and 60%.

## 2.2. Elastic energy

The elastic energy functional is given by

$$E^{el} = \int d\vec{r} \left[\frac{1}{2} C_{ijkl}(\epsilon_{ij} - \epsilon_{ij}^T)(\epsilon_{kl} - \epsilon_{kl}^T)\right] \tag{3}$$

where $C_{ijkl}$ is the elastic modulus field, $\epsilon_{kl}^T$ is the transformation strain field and $\epsilon_{ij}$ is the total strain field. Both the $C_{ijkl}$ and $\epsilon_{kl}^T$ are functions of concentration $c(\vec{r}, t)$:

$$C_{ijkl}(c) = C_{ijkl}^\beta \times (1 - c) + C_{ijkl}^{\beta'} \times (c) \tag{4}$$

$$\epsilon_{ij}^T = \epsilon^{oo} \, c \delta_{ij} \tag{5}$$

where $\epsilon^{oo} = \frac{1}{a}\frac{da}{dc}$, $a$ is the lattice parameter, $C_{ijkl}^\beta$ is the modulus of the disordered phase and $C_{ijkl}^{\beta'}$ is the modulus of the ordered phase. The modulus tensor has cubic symmetry with Zener ratio $A_z = 3.0$, i.e., <001> directions have the lowest elastic modulus. The total strain field $\epsilon(\vec{r})$ is determined by the mechanical equilibrium equation: $\nabla \cdot \sigma = 0$. When the modulus of the system is homogeneous ($C_{ijkl}^\beta = C_{ijkl}^{\beta'}$), the elastic energy $E^{el}$ can be directly written as an explicit functional of the order parameters as derived by Khachaturyan et. al. [46, 47] using the Green's function solution in the reciprocal space. This closed form expression of $E^{el}$ greatly increases the numerical efficiency of the model. When the modulus of the system is inhomogeneous, an iterative procedure [48, 49] is used to compute the strain field. The modulus mismatch is quantified in this work using $R_{elas} = C_{ijkl}^{\beta'}/C_{ijkl}^\beta$. In addition to homogeneous modulus case ($R_{elas} = 1.0$), we simulated inhomogeneous modulus cases with $R_{elas} = 1.4$ and $R_{elas} = 0.6$.

## 2.3. Governing equations

The concentration field, $c(\vec{r}, t)$, and the structural order parameter field, $\eta(\vec{r}, t)$, are evolved with time using the Cahn-Hilliard [50] and Allen-Cahn [51] equations,

$$\frac{\partial c}{\partial t} = \nabla \left( M \nabla \left( \frac{\delta F}{\delta c} \right) \right) + \zeta_c \qquad (6)$$

$$\frac{\partial \eta}{\partial t} = -L \frac{\delta F}{\delta \eta} + \zeta_\eta \qquad (7)$$

where $M$ is the chemical mobility, $L$ is the kinetic coefficient characterizing the order↔disorder transition, $\zeta_c$ and $\zeta_\eta$ are the Langevin-noise terms for $c(\vec{r}, t)$ and $\eta(\vec{r}, t)$, respectively. The governing equations (Eq. 6 and Eq. 7) are solved using the semi-implicit spectral method [52]. When the disordered or ordered phase is metastable, the congruent disorder ↔ order transition occurs by nucleation, which is simulated by the Langevin noise term $\zeta_\eta$ in the Allen-Cahn equation (Eq. 7) [53, 54]. For simplicity, the Langevin noise for concentration is chosen to be zero.

## 2.4. Numerical Implementation

The phase-field model was solved numerically through an in-house GPU code. All the numerical values of the non-dimensional model parameters are listed in Table 1. All the simulations are performed in two dimensions with a system size of $512 \times 512$. The Langevin noise is applied for the duration of $T^{Langevin} = 600$ at the beginning of the simulation. The random number seed for the Langevin noise is kept the same for all the simulations, so that the Langevin noise is identical in all the simulations. This eradicates the statistical deviation in the spinodal microstructures between different alloys, so changes observed in the microstructures are due solely to the change in the model parameters. The microstructures were evolved for a total time duration of $T^{total} = 5000$.

To quantitively analyze the topologically connected phase and discreteness of the microstructure, we defined two parameters, $\theta$ and $\epsilon$, respectively. We constructed a $3 \times 3$ supercell of the simulated microstructures (based on $c(\vec{r}, t)$) and counted the number of solute-rich clusters ($c > 0.5$) and solute-lean clusters ($c < 0.5$). Hoshen-Kopelman algorithm [55] with von Neumann neighborhood was used to count the clusters. The parameters $\theta$ (connectivity parameter) and $\epsilon$ (discreteness parameter) are defined as

$$\theta = \frac{N_B - N_R}{N_B + N_R}; \quad \epsilon = |N_B - N_R| \qquad (8)$$

where $N_B$ is the number of solute-lean clusters (which appear blue in the microstructures) and $N_R$ is the number of solute-rich clusters (which appear red in the microstructures). The parameter $\theta$ lies between $-1$ (solute-lean disordered phase being the matrix) and 1 (solute-rich ordered phase being the matrix). Higher $\epsilon$ values imply higher discreteness of the microstructure. The initial microstructure consists of homogeneous $\beta$ phase in all the simulations, corresponding to the high temperature state.

To classify objectively the microstructures as bi-continuous, solute-rich $\beta'$ (ordered phase) matrix and solute-lean $\beta$ (disordered phase) matrix, we defined a cutoff value for $\theta$. If we consider an isolated solute-lean $\beta$ precipitate in a solute-rich matrix ($N_B = 9$ and $N_R = 1$ in the $3 \times 3$ supercell), we get $\theta = (9 - 1)/(9 + 1) = 0.8$. Hence, we use $\theta = 0.8$ as a cutoff to define solute-rich $\beta'$ matrix, i.e., $\theta \geq 0.8$ indicates solute-rich $\beta'$ matrix. Similarly, $\theta \leq -0.8$ indicates solute-lean $\beta$ matrix. The bi-continuous microstructures have $-0.8 < \theta < 0.8$.

## 3. Results

### 3.1. Alloys Investigated

As we simulate microstructural evolutions in many different alloys in this parametric study, we used a 6-digit naming convention to designate these alloys (MPEAXXYY-ZZ). The first pair of digits (XX) represent the free energy surface, the second pair of digits (YY) represent the equilibrium volume fraction of the ordered $\beta'$ phase determined by the alloy composition, and the third pair of digits represent the modulus mismatch conditions. The four free energy surfaces XX={11,12,21,22} used in the current work are shown in Figure 2, which are referred to as MPEA11, MPEA12, MPEA21 and MPEA22, respectively. The free energy surfaces MPEA11 and MPEA12 both have a spinodal region in the $\beta'$ phase and the microstructural evolution will follow PTP(1). However, the spinodal boundaries and the critical point position of $\beta'$ are different between MPEA11 and MPEA12. The free energy surfaces MPEA21 and MPEA22 both have a spinodal region in the disordered phase $\beta$ and the microstructural evolution will follow PTP(2), but the spinodal boundaries and the critical point position of the $\beta$ phase are different. Note that all the four free energy surfaces have identical global minima (obtained by the common tangent construction) at coordinates $(c, \eta) = (0,0)$, i.e., the disordered solute-lean $\beta$ phase, and $(c, \eta) = (1,1)$, i.e., the ordered solute-rich $\beta'$ phase. The spinodal boundaries, the critical point, and the common tangent points for all the free surfaces used in our work are listed in Table 2.

Three volume fractions of the $\beta'$ phase, i.e., YY={40%,50%,60%}, are considered, which correspond to alloy composition c=0.4, 0.5, 0.6, respectively. Furthermore, three different elasticity conditions are considered: (i) Homogeneous modulus (ZZ=01), i.e., we assume that the elastic moduli of the disordered ($\beta$) and ordered ($\beta'$) phases are identical. (ii) Stiffer (hard) $\beta'$ (ZZ=02), e.g., $C^{\beta'}_{ijkl} = 1.4 C^{\beta}_{ijkl}$ ($R_{elas} = 1.4$). (iii) More compliant (soft) $\beta'$, e.g., $C^{\beta'}_{ijkl} = 0.6 C^{\beta}_{ijkl}$ ($R_{elas} = 0.6$). In the simulations, the modulus mismatch is introduced by varying the modulus of the $\beta'$ phase while keeping the $\beta$ phase modulus constant. The lattice misfit between the $\beta'$ and $\beta$ phases is kept constant ($\epsilon^{oo} = 0.2\%$).

Thus, a total of 36 different alloys are considered in this parametric study to quantify the effects of the free energy surface (MPEA11, MPEA12, MPEA21 and MPEA22), volume fraction (40%, 50% and 60%) and elastic modulus misfit (referred as "Homo", "Hard $\beta'$" and "Soft $\beta'$" hereafter), as shown in Figure 4.

### 3.2. Microstructural evolution along PTP(1) and PTP(2)

The simulated microstructural evolutions for alloys MPEA1150-01 and MPEA1250-01 are shown in Figure 5, which follow PTP(1). The microstructural evolution shown in Figure 6 for the alloys MPEA2150-01 and MPEA2250-01 follow PTP(2). In all these cases, the initial microstructure is identical, i.e., homogeneous $\beta$ phase with $c = 0.5$, and all microstructures are simulated under "Homo" ($R_{elas} = 1.0$) elasticity condition.

Let us consider the microstructural evolutions of MPEA1150-01 and MPEA1250-01 shown in Figure 5. The alloys exhibit congruent ordering, and the ordering transformation occurs through nucleation (assisted by the Langevin noise $\zeta_\eta$) and growth. The microstructure consists of an homogeneous $\beta'$ phase at $t = 0.5$ without any concentration modulations, confirming that the congruent ordering occurs first. The concentration modulations develop later at $t = 120$ and the solute-lean regions undergo disordering so that both the $\beta$ and $\beta'$ phases can be seen in the microstructure. At $t = 120$, the disordering is nearly complete for MPEA1150-01 (Figure 5(b4)) in comparison to MPEA1250-01(Figure 5(d4)). At a later time $t = 1000$, disordering is complete in both the alloys, MPEA1150-01 and MPEA1250-01. The microstructure subsequently coarsens in MPEA1150-01 by $t = 5000$ and a periodic array of solute-lean $\beta$ particles are embedded in the solute-rich $\beta'$ matrix (Figure 5(b6)). However, in MPEA1250-01, the solute-rich $\beta'$ particles, though not as discrete as those seen in MPEA1150-01, are embedded in solute-lean $\beta$ matrix. *Although in both alloys an homogeneous modulus is assumed, and the volume fractions of the two co-existing phases are 50%, there is a clear preference for which phase is the matrix phase in the microstructure based on the free energy surface, and the discontinuous phase could be highly discrete*. Note that under similar conditions (homogeneous modulus and 50% volume fraction), a bi-continuous microstructure will be observed in a binary alloy with a symmetrical miscibility gap (i.e., just an isostructural transformation without congruent ordering/disordering) [56].

In Figure 6, the microstructural evolution in alloys MPEA2150-01 and MPEA2250-01 follow PTP(2), i.e., the alloys do not undergo congruent ordering and the microstructures are composed of only $\beta$ phase without any concentration modulation as observed at $t = 0.5$ in contrast to the alloys in Figure 5. The concentration modulations develop at $t = 120$. Along the concentration modulations, an ordering transformation occurs in the solute-rich regions of both alloys and the ordering is nearly complete in MPEA2150-01 at $t = 120$, while MPEA2250-01 is still in its early stage of ordering at the same time. At later times (e.g., $t = 5000$), a microstructure with a periodic array of discrete solute-rich $\beta'$ particles embedded in the solute-lean $\beta$ matrix is observed in case of MPEA2150-01, while a bi-continuous $\beta'/\beta$ microstructure is observed in MPEA2250-01, although both alloys followed the same transformation pathway PTP(2). This difference in microstructure is due solely to the difference in the free energy surface.

*3.3. Simulated microstructures for all the alloys*

As expected, we observed that all the alloys having the free energy surface MPEA11 and MPEA12 follow PTP(1), while the alloys having the free energy surface MPEA21 and MPEA22 follow PTP(2). For brevity, we have not shown the microstructural evolution of all the alloys in this article. Instead, the final microstructures represented by $c(\vec{r})$ of all alloys are grouped together based on the volume fraction ($f_V$) and are presented in Figure 7 ($f_V = 40\%$), Figure 8 ($f_V = 50\%$) and Figure 9 ($f_V = 60\%$), respectively. Note that the volume fraction always refers to the solute-rich $\beta'$ phase unless indicated otherwise. To make an objective comparison between the observed microstructures, we calculate the connectivity $\theta$ and the discreteness $\epsilon$ of all the alloys and these are shown in Figure 10 and Figure 11, respectively. In addition, the final volume fractions in all the alloys were verified to be within 1% of the equilibrium volume fraction (Supplementary Material: Section A). All the alloys with 40% volume fraction (Figure 7) exhibited $\beta$ matrix + $\beta'$ precipitate except for MPEA1140-03 (Figure7(a3)) that exhibits a bi-continuous microstructure. However, the discreteness $\epsilon$ of the microstructures drastically varies depending on the free energy surface and modulus mismatch conditions as can be observed in Figure 11(a). The alloy MPEA2140-02 (Figure 7(c2)) has the maximum precipitate discreteness. Similarly, all the alloys with 60% volume fraction (Figure 9) exhibit $\beta'$ matrix + $\beta$ precipitate microstructures, except for MPEA2160-02 (Figure 9(c2), which shows a bi-continuous microstructure. In this case, the maximum discreteness is observed for MPEA1160-03 (Figure 9(a3)). In alloys with $f_V = 50\%$ (Figure 8), microstructures with both $\beta$ matrix + $\beta'$ precipitates and $\beta'$ matrix + $\beta$ precipitates are observed along with a bi-continuous microstructure corresponding to MPEA2250-01 (Figure 8(d1)).

## 4. Discussion

*4.1. Phase Transformation Pathways (PTPs)*

The microstructural evolutions shown in Figure 5 and Figure 6 demonstrate the phase transformation pathways PTP(1) and PTP(2), respectively. The as-cast microstructures of bcc/B2 HEAs have exhibited (i) $\beta$ matrix + $\beta'$ precipitate (ii) $\beta'$ matrix + $\beta$ precipitate (iii) bi-continuous $\beta/\beta'$ microstructures [42, 57-61]. The simulated microstructures have shown all these morphologies through PTP(1) and PTP(2) as quantified through $\theta$ in Figure 10. If the phase decomposition occurs through nucleation and growth alone, we expect only $\beta$ matrix + $\beta'$ precipitates. The simulated microstructures highlight the versatility of the PTP(1) and PTP(2) in obtaining all the various microstructures observed in the experiments.

The evolutions in Figure 5 and Figure 6 clearly highlight the coupling between the order ↔ disorder transition and spinodal decomposition processes leading to a mixture of ordered and disordered phases. Recently, Li et al. [42] proposed the classical spinodal decomposition process as a possible PTP to explain the B2 + bcc microstructures observed in Al-Ni-Co-Fe-Cr system. Although the simulated microstructures are similar to the experimentally observed ones, the proposed PTP does not consider the order ↔ disorder transition.

Spinodal decomposition (long-range diffusion) is a slow process compared with congruent ordering (short-range diffusion). This is evident from the difference in timescales of the congruent ordering ($t = 0.5$) and spinodal decomposition ($t = 120$) observed in both alloys in Figure 5. Also, the disordering of the solute-lean region in MPEA1250-01 (Figure 5(d4)) is slower compared with that in MPEA1150-01 (Figure 5(b4)). This slow disordering is expected because the intersection of the ordered and disordered free energy curve ($c = 0.2$) is further away from the alloy composition $c = 0.5$ in the case of MPEA12 compared with the intersection point $c = 0.3$ in MPEA11 (Figure 3). The concentration modulations should go beyond the intersection point so that the disordering process is energetically feasible (i.e., disordering would decrease the free energy). Thus, in MPEA12, the decomposition should proceed significantly longer before the start of the disordering process, which leads to a much delayed disordering process in MPEA1250-01. Similarly, the ordering of the solute-rich region in MPEA2250-01 (Figure 6(d4)) is slower compared with that in MPEA2150-01 (Figure 6(b4)), because the intersection point in MPEA22 ($c = 0.8$) is further away from the alloy composition $c = 0.5$ compared with the intersection point of MPEA21 ($c = 0.7$) as shown in Figure 3.

*4.2. Effect of an asymmetry factor ($A_s$)*

We hypothesize that the location of the critical point ($c_m$) (refer Figure 3) with respect to the equilibrium compositions of the final two equilibrium phases (referred as $c_\beta^{eq}$ and $c_{\beta'}^{eq}$) plays an important role in determining the morphology of the microstructure in addition to the well-known effect of modulus mismatch ($R_{elas}$) and equilibrium volume fraction ($f_V$) [43, 45, 56]. To quantify the relative location of $c_m$, we define an asymmetry factor $A_s$ as,

$$A_s = \frac{2c_m}{c_{\beta'}^{eq} + c_\beta^{eq}}. \tag{9}$$

In a symmetric system, where the critical point is located at the mid-point of the equilibrium tie-line, we have $A_s = 1.0$. If the critical point is shifted either to the right or to the left, the asymmetry factor would become either $A_s > 1.0$ or $A_s < 1.0$, respectively. For instance, $A_s$ values for the free energy surfaces of MPEA11, MPEA12, MPEA21 and MPEA22 are listed in Table 2. It is quite clear that all of them are asymmetrical systems. In these asymmetrical systems, if $A_s < 1$ ($A_s > 1$), the volume fraction of the ordered (disordered) phase at the early stage of decomposition is larger than the other phase, which in turn would lead to an ordered (disordered) matrix phase. However, at the later stages, the volume fractions of the individual phases would reach the ones predicted by the lever rule with the equilibrium tie-line.

In order to isolate the effect of $A_s$ from that of the modulus mismatch ($R_{elas}$) and volume fraction ($f_V$), we focus on the alloys with $R_{elas} = 1.0$ and $f_V = 50\%$, i.e., MPEA1150-01, MPEA1250-01, MPEA2150-01 and MPEA2250-01. In these alloys, neither the elastic modulus nor the equilibrium volume fraction prefers a $\beta$ matrix or a $\beta'$ matrix. The calculated $\theta$ and $A_s$ in these alloys are plotted against each other in Figure 12 along with the microstructures. It shows a strong correlation between $\theta$ and $A_s$. As mentioned above, if $A_s < 1$ the free energy surface clearly prefers $\beta'$ to be the matrix and if $A_s > 1$, the free energy surface prefers $\beta$ to be the matrix. This preference increases with $|A_s - 1|$, which can be regarded as the degree of asymmetry of the free energy surface. Although for both MPEA1150-01 and MPEA2250-01, we have $A_s < 1.0$, only MPEA1150-01 exhibits a $\beta'$ matrix whereas MPEA2250-01 has a bi-continuous microstructure. It is because $|1 - A_s|$ is larger for MPEA1150-01 compared with MPEA2250-01.

Similarly, both MPEA1250-01 ($A_s = 1.10$) and MPEA2150-01 ($A_s = 1.18$) exhibit a $\beta$ matrix, although the phase transformation pathway is different for these two alloys (PTP(1) and PTP(2), respectively). In MPEA1250-01, the $\beta'$ phase undergoes spinodal decomposition, whereas in MPEA2150-01 the $\beta$ phase undergoes spinodal decomposition. As both alloys have similar $A_s > 1$ which led to similar connectivity $\theta$ of the microstructure, we can expect that $A_s$ has a stronger effect on the microstructure than that of the PTP. This understanding could allow one to focus on just a miscibility gap $\beta_1 + \beta_2$ without considering the ordering reaction and explore just the effect of the asymmetry factor $A_s$. This could allow us to study alloys with much larger $|A_s - 1|$, which would be a focus of future work.

In spinodal systems, we expect bi-continuous microstructures for $f_V = 50\%$ if there is no modulus mismatch between the phases [56]. In the existing literature of multi-phase HEAs, only modulus mismatch is discussed as a viable mechanism to obtain a unique matrix phase and to obtain discrete precipitates [22, 32, 42, 61]. In this study, we have shown for the first time that highly discrete precipitates embedded in a continuous matrix can be achieved by spinodal decomposition in a binary miscibility gap with equal volume fractions of the two phases (i.e., $f_V = 50\%$). The discrete precipitate microstructures obtained resemble those observed in Ni-base superalloys formed by nucleation and growth [62-64].

*4.3. Competing effects between volume fraction ($f_V$) and asymmetry factor ($A_s$)*

The parameters $\theta$ and $\epsilon$ as a function of $A_s$ for all the elasticity conditions are shown in Figure 13 and Figure 14, respectively. In this section, we will discuss only $R_{elas} = 1.0$ (Homo) alloys to isolate the effect of volume fraction and asymmetry of the free energy, i.e. alloys shown in Figure 13(a) and Figure 14(a).

In typical homogeneous modulus systems, the majority phase (i.e., the phase with higher volume fraction $f_V > 50\%$) prefers to be the matrix and the minority phase (i.e., the phase with lower volume fraction $f_V < 50\%$) exists as embedded particles [56]. This effect is observed in our alloys as shown in Figure 13(a), where for $f_V = 60\%$, the $\beta'$ phase is the matrix and for $f_V = 40\%$, the $\beta$ phase is the matrix. Thus, in both cases, the majority phase is the matrix phase, independent of the value of $A_s$.

However, the discreteness ($\epsilon$) of the microstructures varies significantly with $A_s$ (Figure 14(a)). For $f_V = 40\%$, $\epsilon$ is the smallest for $A_s = 0.82$ (MPEA1140-01, Figure 7(a1)), and largest for $A_s = 1.18$ (MPEA2140-01, Figure 7(c1)). In the case of $A_s = 0.82$, the asymmetry of the free energy prefers a $\beta'$ matrix ($A_s < 1.0$), whereas the actual microstructure has a $\beta$ matrix due to the bias from volume fraction ($f_V < 50\%$). These opposing factors lead to a lower $\epsilon$. In contrast, for $A_s = 1.18$, both the asymmetry ($A_s > 1$) and volume fraction ($f < 50\%$) prefer $\beta$ matrix, leading to larger value of $\epsilon$. Similarly, the discreteness trend observed for $f_V = 60\%$ can also be explained. In this case, $\epsilon$ is the largest for $A_s = 0.82$ (MPEA1160-01, Figure 9(a1)) because both asymmetry ($A_s < 1$) and volume fraction ($f_V > 50\%$) prefer a $\beta'$ matrix.

*4.4. Competing effects between modulus mismatch ($R_{elas}$) and asymmetry factor ($A_s$)*

The effect of modulus mismatch on the microstructural evolution during spinodal decomposition was first studied by Onuki and Nishimori [43, 65]. They showed that the modulus difference between the co-existing phases during phase separation can significantly alter the microstructure. As per their calculations, the elastically more compliant phase becomes the matrix and the elastically stiffer phase becomes the precipitate. Further, the increase in the modulus of the precipitate can significantly lower the coarsening rate of the precipitates resulting in highly discrete microstructures [43]. Thus, in the case of $R_{elas} = 1.4$ (Hard $\beta'$), the modulus mismatch prefers the $\beta$ matrix, and in the case of $R_{elas} = 0.6$ (Soft $\beta'$), the modulus mismatch prefers the $\beta'$ matrix. In this subsection, we will only discuss the alloys with $f_V = 50\%$ (Figure 10b), so that the volume fraction ($f_V$) does not influence the topology of the microstructure.

For $f_V = 50\%$, the free energy surface MPEA22 leads to a $\beta$ matrix for $R_{elas} = 1.4$ (Figure 8(d2)), although the asymmetry factor favors a $\beta'$ matrix in MPEA22 ($A_s = 0.88$). In this case, the $\beta$ phase becomes the matrix phase because of the dominant modulus mismatch effect. Similarly, in MPEA12, we observe $\beta'$ as the matrix for $R_{elas} = 0.6$ (Figure 8(b3)), even though the asymmetry factor favors a $\beta$ matrix ($A_s = 1.1$). The time evolution of $\theta$ is plotted in Figure 15 for MPEA2250-02 and MPEA1250-03. It is readily seen that $\theta$ changes sign during the evolution for both alloys. MPEA2250-02 had a $\beta'$ matrix initially (preferred by asymmetry) and upon coarsening, the microstructure is inverted to a $\beta$ matrix (preferred by the modulus mismatch). Similarly, the microstructure of MPEA1250-03 changed its connectivity from a $\beta$ matrix to a $\beta'$ matrix. This phenomenon is termed "phase inversion" [44]. Note that if the miscibility gap is symmetric (i.e., $A_s = 1.0$), when the volume fraction is $f_V = 50\%$, the microstructure attains the matrix phase preferred by the modulus mismatch [43, 65] and phase inversion would not occur. However, in our case $A_s \neq 1$ and the initial connectivity $\theta$ is based on the asymmetry of the free energy ($A_s$) rather than the modulus mismatch. During coarsening, the contribution from the elastic energy increases and it becomes energetically favorable to have the low modulus phase as the matrix phase which causes the phase inversion.

Interestingly, the free energy surface MPEA11 leads to a $\beta'$ matrix (preferred by $A_s = 0.82$) even for the case $R_{elas} = 1.4$ where the modulus mismatch prefers a $\beta$ matrix (Figure 8(a2) and Figure 10(b)). This suggests that the effect of the asymmetry factor is dominant over the modulus mismatch in MPEA11 compared with MPEA22. The dominant asymmetry effect in MPEA11 is likely due to the larger $|A_s - 1|$ value in MPEA11 compared with that in MPEA22. However, for a long enough aging time, the modulus mismatch effect would dominate due to the increase in precipitate size. We performed a long-time aging simulation ($T^{total} = 50,000$) for MPEA1150-02 to check whether the phase inversion would occur. The comparison between the connectivity $\theta$ for MPEA1150-02 and MPEA2250-02 is shown in Figure 16. Phase inversion indeed happens at a longer aging time for MPEA1150-02. However, the time taken for the phase inversion is much longer due to the stronger negative contribution of the asymmetry of the free energy $|A_s - 1|$ in MPEA11. Note that our analysis does not consider the loss of coherency which may occur during long-time aging that can further impact the phase inversion behavior.

The discreteness $\epsilon$ of the alloys with $f_V = 50\%$ are shown in Figure 11(b). In the case of $R_{elas} = 1.4$ (Hard $\beta'$), both the free energy surfaces MPEA11 ($A_s = 0.82$) and MPEA22 ($A_s = 0.88$) lead to smaller $\epsilon$. This is because the asymmetry factor ($A_s < 1$) favors a $\beta'$ matrix and the modulus mismatch ($R_{elas} > 1$) favors a $\beta$ matrix. These opposing effects lead to a smaller $\epsilon$ (less discrete) of the microstructures. Similarly, the free energy surfaces MPEA12 and MPEA21 both lead to larger $\epsilon$ compared with those associated with the free energy surfaces MPEA11 and MPEA22. This is because in MPEA12 and MPEA21 both the asymmetry ($A_s > 1$) and the modulus mismatch ($R_{elas} > 1$) favor a $\beta$ matrix. Also, the free energy surface MPEA21 leads to a larger $\epsilon$ compared with that associated with MPEA12, because of the larger $|A_s - 1|$ in MPEA21 compared to with that in MPEA12.

*4.5. Competing effects between asymmetry factor ($A_s$), volume fraction ($f_V$) and modulus mismatch ($R_{elas}$)*

In this subsection, we compare the microstructures of all the alloys in terms of the calculated $\theta$ and $\epsilon$, to establish the interplay among $A_s$, $f_V$ and $R_{elas}$. The preferences of these factors to either a $\beta$ matrix or a $\beta'$ matrix are summarized in Table 3. The calculated $\theta$ and $\epsilon$ of the final microstructures as a function of $A_s$ for all the elasticity conditions are shown in Figure 13 and Figure 14, respectively.

### 4.5.1. Bi-continuous microstructures

Bi-continuous microstructures are obtained when there is no preference of matrix phase for any of the factors or when there are opposing factors preferring different matrix phases. Based on the connectivity $\theta$ (Figure 13), we identified three bi-continuous microstructures in our simulations: (i) MPEA1140-03 (Figure 7(a3)), (ii) MPEA2160-02 (Figure 9(c2)) and (iii) MPEA2250-01 (Figure 8(d1)).

- MPEA1140-03 ($A_s = 0.82$, $f = 40\%$, $R_{elas} = 0.6$): In this alloy, the volume fraction prefers a $\beta$ matrix while the asymmetry factor and modulus mismatch prefer a $\beta'$ matrix. This competition led to a bi-continuous microstructure. However, upon prolonged annealing we would expect $\beta'$ evolving into the matrix as the modulus misfit becomes dominant in the later stages of coarsening [43, 44].
- MPEA2160-02 ($A_s = 1.18$, $f = 60\%$, $R_{elas} = 1.4$): In this alloy, the volume fraction prefers $\beta'$ matrix while the asymmetry factor and modulus mismatch prefer a $\beta$ matrix, which led to a bi-continuous microstructure. Here, prolonged annealing would lead to a $\beta$ matrix.
- MPEA2250-01 ($A_s = 0.88$, $f = 50\%$, $R_{elas} = 1.0$): Here, both the volume fraction and the modulus mismatch exhibit no preference, only the asymmetry factor prefers a $\beta'$ matrix. However, as the strength of preference $|A_s - 1|$ is relatively small, the microstructure is bi-continuous. In this case, even after prolonged annealing we would still expect a bi-continuous microstructure.

### 4.5.2. Discreteness of the microstructures

- $\beta'$ matrix + $\beta$ precipitate: As observed from Figure 13, most alloys with $f_V = 60\%$ and few ones with $f_V = 50\%$ exhibit $\beta'$ matrix + $\beta$ precipitate. No alloy with $f_V = 40\%$ exhibits a $\beta'$ matrix, suggesting the strong impact of the volume fraction on the microstructure in comparison to the other parameters such as $A_s$ and $R_{elas}$. However, upon increasing the strength of the asymmetry $|A_s - 1|$, it may be possible to obtain a $\beta'$ matrix even for $f_V = 40\%$, which would be explored in our future work. Among the microstructures having $\beta'$ matrix, the maximum discreteness is observed for $A_s = 0.82$ in $f_V = 60\%$ with $R_{elas} = 0.6$ (Figure 14(c)). The corresponding microstructure (MPEA1160-03) is shown in Figure 9(a3). This microstructure is highly discrete because the asymmetry ($A_s = 0.82$), volume fraction ($f_V = 60\%$) and modulus mismatch ($R_{elas} = 0.6$) synergistically favor a $\beta'$ matrix.
- β matrix + β′ precipitate: Most alloys with $f_V = 40\%$ and few alloys with $f_V = 50\%$ exhibit a $\beta$ matrix. The maximum discreteness is observed in MPEA2140-02 ($A_s = 1.18$, $f = 40\%$, $R_{elas} = 1.4$) as shown in Figure 14(b) and the microstructure is shown in Figure 7(c2). This is because all these factors favor a $\beta$ matrix.

### 4.6. Limitations

During the congruent ordering process shown in Figure 5(b1)-(b3), anti-phase domains (APDs) can form (e.g.. two APDs for bcc → $B2$ transformation and 4 APDs for fcc → $L1_2$ transformation. The effects of APDs and antiphase domain boundaries (APBs) on microstructural evolution have been well documented by phase-field simulations in the literature [38, 66]. For example, when congruent ordering precedes decomposition (such as in PTP(1)), the disordered phase preferably forms at APBs during decomposition [38]. The presence of APDs also prevents particle coalescence during growth and coarsening of the ordered precipitates, increasing the discreteness of the precipitates [67]. In the current study, the effect of APBs is not considered as we would like to isolate the effect of other factors taken into account. When the ordered phase occurs as the matrix phase, our analysis represents cases where the sizes of the ordered domains are much larger the precipitate size, so the effect of APBs could be ignored. For instance, in AlMo$_{0.5}$NbTa$_{0.5}$TiZr where B2 is the matrix phase, the size of the ordered domains is much larger compared with the precipitate size [16, 33, 45].

The elastic moduli of the ordered and disordered phase, and the lattice misfit are coupled only with concentration but not the structural order parameter, which means that during the congruent ordering (or disordering), the elastic moduli and lattice misfit does not change. This is obviously not the case in experiments and order↔disorder transtition can induce sudden change in modulus mismatch and lattice misfit which can cause morphological changes in the microstructure (e.g. particle splitting [56, 68, 69]). This effect should be considered in future investigations.

## 5. Conclusion

In this work, we have simulated microstructural evolution along two different phase transformation pathways (PTPs): (1) (i) congruent ordering followed by spinodal decomposition in the ordered phase and then disordering of one of the ordered phases, i.e., $\beta \rightarrow \beta' \rightarrow \beta'_1 + \beta'_2 \rightarrow \beta + \beta'_2$ and (2) spinodal decomposition in the disordered phase followed by ordering of one of the disordered phases, i.e., $\beta \rightarrow \beta_1 + \beta_2 \rightarrow \beta_1 + \beta'$, where $\beta'$ represents an ordered phase and $\beta$ represents a disordered phase. Using high throughput phase field simulations of 36 alloys, we identified the key parameters and their interplays that affect the microstructural characteristics developed along these PTPs, including $\beta'$ matrix + $\beta$ precipitates, $\beta$ matrix + $\beta'$ precipitates, and the discreteness of the precipitate phase. These factors include (a) the equilibrium volume fraction of the ordered phase ($f_V$), (b) the asymmetry factor of the free energy surface (i.e., the location of the critical point of the miscibility gap relative to the compositions of the final two equilibrium phases) and the modulus mismatch between the ordered and disordered phases ($R_{elas}$). The microstructures are quantified through a connectivity parameter, $\theta$, and a discrete parameter, $\epsilon$. The following are the key observations from the parametric study:

1. The location of the critical point ($c_m$) with respect to the equilibrium tie-line plays a vital role in the microstructural evolution. We quantified this effect using an asymmetry factor, $A_s$. This effect is previously ignored in the literature. With this effect, we showed that highly discrete precipitates embedded in a continuous matrix could be achieved by spinodal decomposition with equal volume fractions and equal elastic modulus of the two phases (i.e., $f_V = 50\%$ and $R_{elas} = 1$). The discrete precipitate microstructures obtained resemble those observed in Ni-base superalloys formed by nucleation and growth. Both $\beta$ precipitates + $\beta'$ matrix ($A_s < 1$) and $\beta'$ precipitates + $\beta$ matrix ($A_s > 1$) could be obtained by the appropriate design of the asymmetry parameter ($A_s$).
2. Key parameters that determine which phase is the matrix phase and which phase is the precipitate phase:
    - Volume fraction $f_V$: $f_V > 50\%$ prefers $\beta'$ matrix, $f_V < 50\%$ prefers a $\beta$ matrix.

- Asymmetry $A_s$: $A_s$ determines the volume fractions of the solute-lean and solute-rich phases at the early stages of spinodal decomposition, which could be quite different from the equilibrium volume fractions determined by the tie-line. $A_s < 1$ prefers a $\beta'$ matrix and $A_s > 1$ prefers a $\beta$ matrix.

- Modulus mismatch $R_{elas}$: $R_{elas} < 1$ (soft $\beta'$) prefers a $\beta'$ matrix, $R_{elas} > 1$ (hard $\beta'$) prefers a $\beta$ matrix.

3. Highly discrete microstructure can be obtained by synergistically combining various factors influencing the microstructure:
   - $\beta$ matrix + $\beta'$ precipitates: $A_s > 1$, $f_V < 50\%$, $R_{elas} > 1$ (demonstrated by the alloy MPEA2140-02, Figure 7(c2)).
   - $\beta'$ matrix + $\beta$ precipitates: $A_s < 1$, $f_V > 50\%$, $R_{elas} < 1$ (demonstrated by the alloy MPEA1160-03, Figure 9(a3)).
4. A phase inversion phenomenon could be observed during the spinodal-meditated PTP by the competition between the asymmetry factor of the free energy ($A_s$) and the modulus mismatch ($R_{elas}$).

This systematic, high throughput parametric study will afford help in alloy design of MPEAs with desired microstructures for specific engineering applications.

**Acknowledgements:** The authors KK and YW would like to acknowledge the financial support by Air Force Office of Scientific Research (AFOSR) under grant FA9550-20-1-0015. Computational resources for this work are provided by Ohio Supercomputer Center under Project No. PAS0971. KK thanks Dr. Shalini Roy Koneru for proofreading the manuscript. KK acknowledges fruitful discussions with Professor T.A. Abinandanan at Indian Institute of Science, Bangalore. Author HLF acknowledges the support in part of the Center for the Accelerated Maturation of Materials (CAMM) at The Ohio State Univesity.

**Tables**

Table 1. Model parameters used in the phase-field model.

| Parameter name | Symbol | Value |
|---|---|---|
| Chemical Mobility | M | 1 |
| Interface kinetic coefficient | L | 30 |
| Gradient energy coefficient of $c(\vec{r},t)$ | $\kappa_c$ | 1.00 |
| Gradient energy coefficient of $\eta(\vec{r},t)$ | $\kappa_\eta$ | 0.25 |
| Double-well barrier height | $\omega$ | 0.25 |
| Elastic modulus tensor | $C_{1111}^\beta = C_{11}^\beta$ | 37,777 |
| Elastic modulus tensor | $C_{1122}^\beta = C_{12}^\beta$ | 31,111 |
| Elastic modulus tensor | $C_{2323}^\beta = C_{44}^\beta$ | 10,000 |
| Lattice misfit | $\epsilon^{oo}$ | 0.2% |
| Modulus mismatch | $R_{elas} = \dfrac{C_{ijkl}^{\beta'}}{C_{ijkl}^{\beta}}$ | 1.0 (Homo)<br>1.4 (Hard $\beta'$)<br>0.6 (Soft $\beta'$)<br>(*refer Section 3.1 for details*) |
| Gaussian random number generator with zero mean and unit standard deviation | $\rho$ | - |
| Langevin noise for $\eta(\vec{r},t)$ | $\zeta_\eta$ | $\rho\sqrt{\dfrac{2(k_B T)L}{\Delta t}}$<br>(i) $\Delta t$ – discretization timestep.<br>(ii) $k_B T = 0.01$ non-dimensional thermal energy. |
| Discretization timestep | $\Delta t$ | $0.001\ (t \leq 0.5)$<br>$0.01\ (0.5 < t \leq 600)$<br>$0.10\ (t > 600)$ |
| Duration of Langevin noise in the beginning of the simulation | $T^{Langevin}$ | 600 |
| Total duration of the simulation | $T^{total}$ | 5000 |

Table 2. Spinodal region, common-tangent points, critical points and the asymmetry factor ($A_s$) of all the free energy surfaces used in this work.

| Free energy surface | Spinodal occurs in phase | Spinodal region ($c_{min}^{sp}, c_{max}^{sp}$) | Critical point ($c_m$) | Common tangent points ($c, \eta$) | Asymmetry factor ($A_s$) |
|---|---|---|---|---|---|
| MPEA11 | $\beta'$ | (0.08, 0.74) | 0.41 | (0,0) and (1,1) | 0.82 |
| MPEA12 | $\beta'$ | (0.31, 0.80) | 0.55 | (0,0) and (1,1) | 1.10 |
| MPEA21 | $\beta$  | (0.26, 0.92) | 0.59 | (0,0) and (1,1) | 1.18 |
| MPEA22 | $\beta$  | (0.20, 0.69) | 0.44 | (0,0) and (1,1) | 0.88 |

Table 3. Parameters affecting the microstructure morphology along with their preference and their degree of preference for different microstructures in both the phase transformation pathways PTP(1) and PTP(2).

| Parameter | $\beta$ matrix + $\beta'$ precipitate | $\beta'$ matrix + $\beta$ precipitate | degree of preference |
|---|---|---|---|
| Asymmetry of the free energy ($A_s$) | $A_s > 1$ | $A_s < 1$ | $|A_s - 1|$ |
| Volume fraction of $\beta'$ phase ($f$) | $f < 0.5$ | $f > 0.5$ | $|f - 0.5|$ |
| Modulus mismatch ($R_{elas} = C_{ijkl}^{\beta'}/C_{ijkl}^{\beta}$) | $R_{elas} > 1$ | $R_{elas} < 1$ | $|R_{elas} - 1|$ |

**Figures**

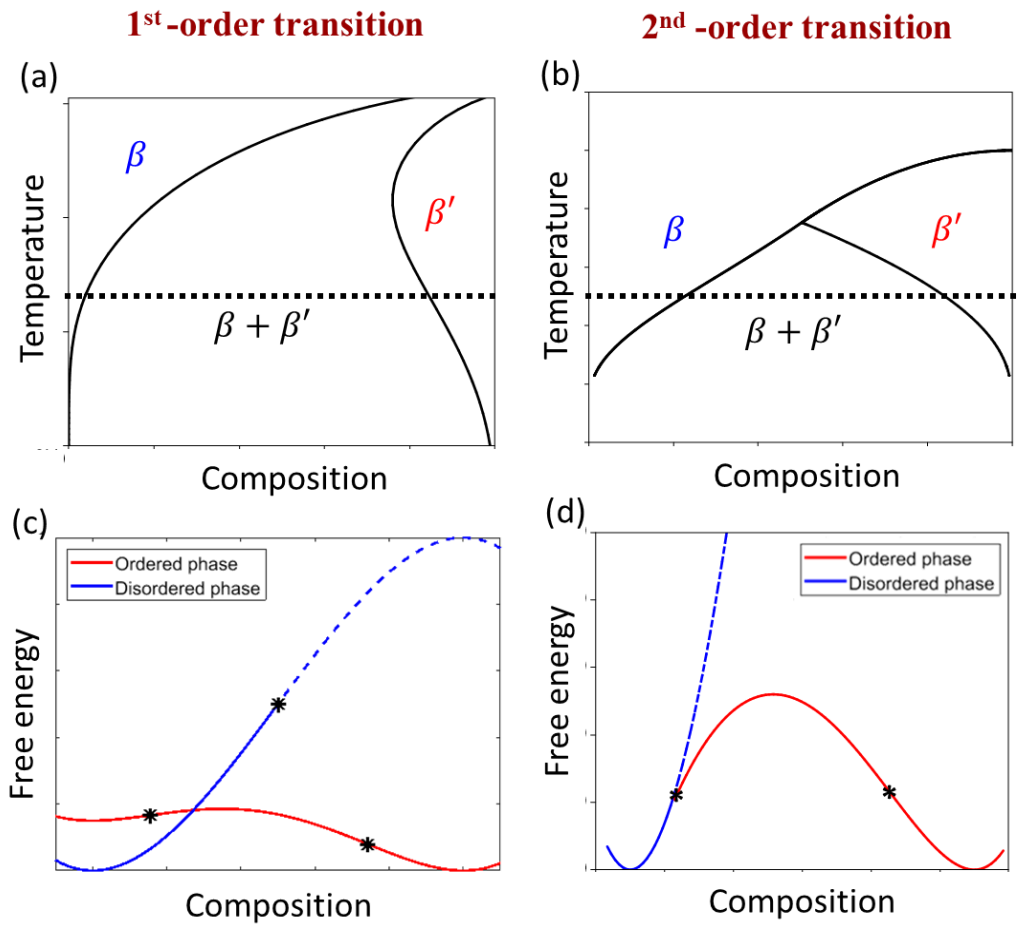

Figure 1. Schematic of the phase diagram and free energy curves. (a) Phase diagram for 1$^{st}$ order transition. (b) Phase diagram for 2$^{nd}$ order transition. (c) Free energy of the ordered and disordered phase at a temperature marked by horizontal dotted line in (a). (d) Free energy of the ordered and disordered phase at the temperature marked by horizontal dotted line in (b). The blue dashed line in (c,d) represent the absolute unstable region. The stars in free energy of ordered phase represents the spinodal boundaries.

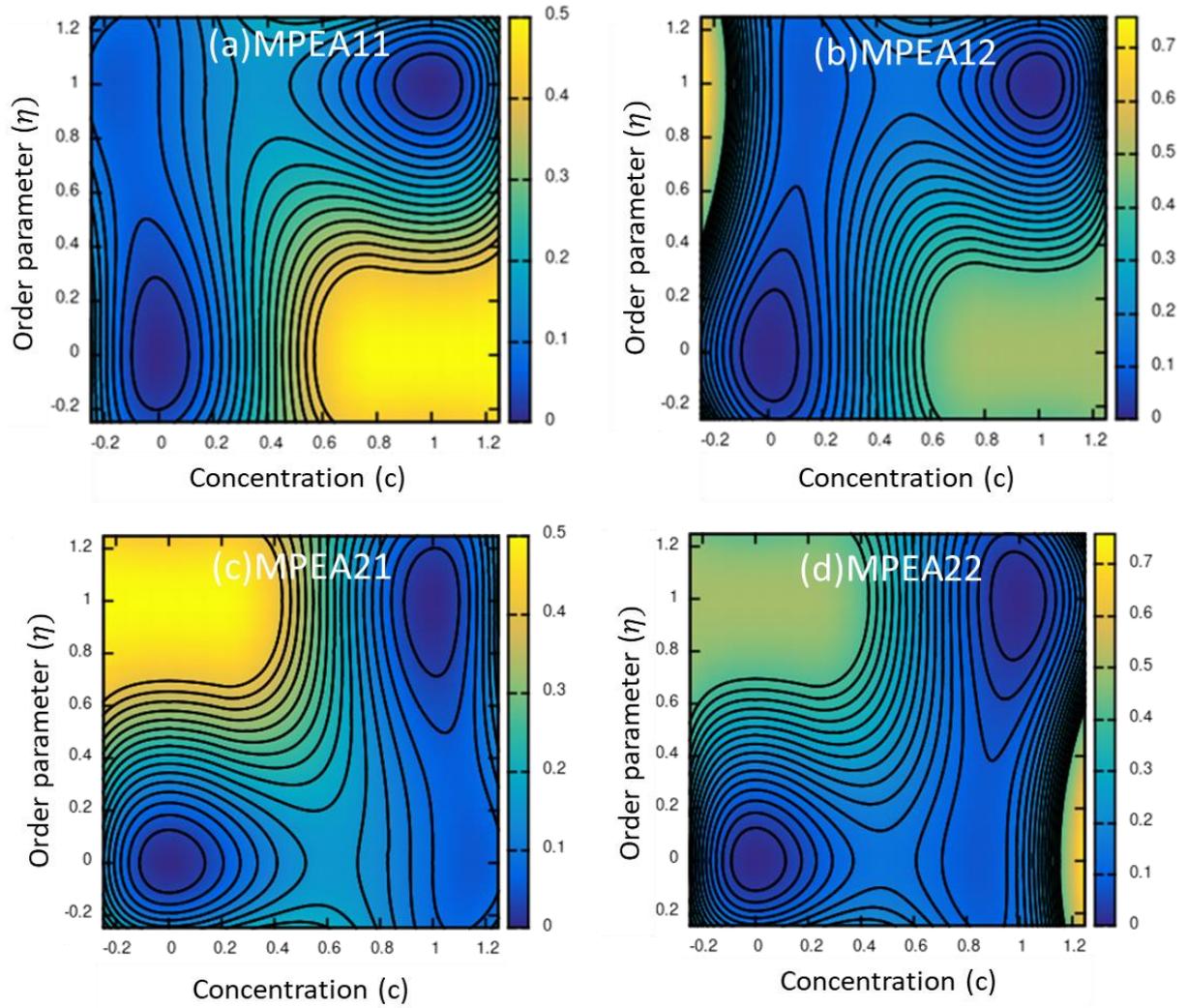

Figure 2. Free energy surfaces $f(c,\eta)$ used in our study as a function of concentration c (x-axis) and order parameter $\eta$ (y-axis). (a) MPEA11 (b) MPEA12 (c) MPEA21 (d) MPEA22. In all surfaces, the common tangent construction gives (0,0) and (1,1) i.e., the solute-lean disordered ($\beta$) and solute-rich ordered ($\beta'$) phases respectively as equilibrium phases.

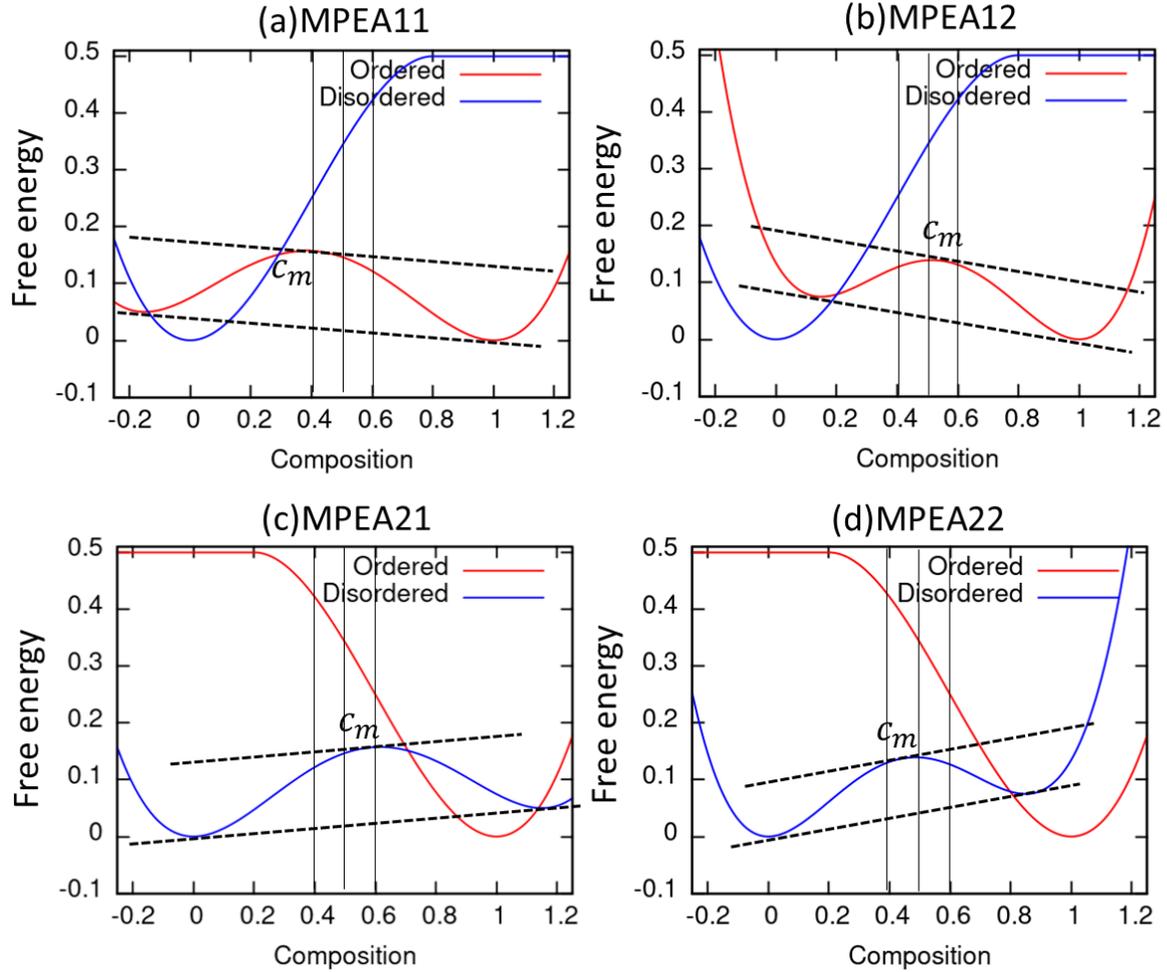

Figure 3: Free energy curves of the ordered ($\eta = 1$) and disordered ($\eta = 0$) phases from the projection of the free energy surfaces in Fig. 2 onto the $f$-$c$ plane. (a) MPEA11 (b) MPEA12 (c) MPEA21 (d) MPEA22. The three vertical lines in each subfigure at $c_{avg} = 0.4, 0.5$ and $0.6$ shows the alloy compositions which will be explored in our work. The dashed lines highlight the critical point ($c_m$) and miscibility gap variation amongst the different free energy surfaces. The exact values of the critical points are listed in Table 2. Note that the non-dimensional solute concentration ($c$) can be related to real solute concentration ($x$) using $c = (x - x_\beta)/(x_{\beta'} - x_\beta)$ where $x_\beta$ and $x_{\beta'}$ are the equilibrium solute concentrations of the $\beta$ and $\beta'$ phases, respectively.

| S.No | Alloy | S.No | Alloy |
|---|---|---|---|
| 1 | MPEA1140-01 | 19 | MPEA2140-01 |
| 2 | MPEA1150-01 | 20 | MPEA2150-01 |
| 3 | MPEA1160-01 | 21 | MPEA2160-01 |
| 4 | MPEA1240-01 | 22 | MPEA2240-01 |
| 5 | MPEA1250-01 | 23 | MPEA2250-01 |
| 6 | MPEA1260-01 | 24 | MPEA2260-01 |
| 7 | MPEA1140-02 | 25 | MPEA2140-02 |
| 8 | MPEA1150-02 | 26 | MPEA2150-02 |
| 9 | MPEA1160-02 | 27 | MPEA2160-02 |
| 10 | MPEA1240-02 | 28 | MPEA2240-02 |
| 11 | MPEA1250-02 | 29 | MPEA2250-02 |
| 12 | MPEA1260-02 | 30 | MPEA2260-02 |
| 13 | MPEA1140-03 | 31 | MPEA2140-03 |
| 14 | MPEA1150-03 | 32 | MPEA2150-03 |
| 15 | MPEA1160-03 | 33 | MPEA2160-03 |
| 16 | MPEA1240-03 | 34 | MPEA2240-03 |
| 17 | MPEA1250-03 | 35 | MPEA2250-03 |
| 18 | MPEA1260-03 | 36 | MPEA2260-03 |

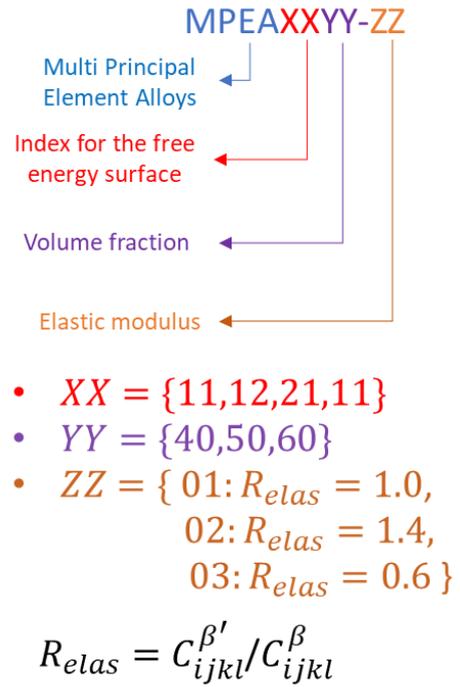

Alloy naming convention:

MPEAXXYY-ZZ

- MPEA: Multi Principal Element Alloys
- XX: Index for the free energy surface
- YY: Volume fraction
- ZZ: Elastic modulus

- $XX = \{11, 12, 21, 11\}$
- $YY = \{40, 50, 60\}$
- $ZZ = \{\, 01: R_{elas} = 1.0,\ 02: R_{elas} = 1.4,\ 03: R_{elas} = 0.6 \,\}$

$$R_{elas} = C_{ijkl}^{\beta'} / C_{ijkl}^{\beta}$$

Figure 4. Simulated alloys in this work. The naming convention (MPEAXXYY-ZZ) is as follows: $XX = \{11,12,21,11\}$ represents the free energy surface. $YY = \{40,50,60\}$ represents the volume fraction of $\beta'$. $ZZ = \{01: R_{elas} = 1.0, 02: R_{elas} = 1.4, 03: R_{elas} = 0.6\}$ represents the elasticity conditions. The elasticity conditions 01, 02, and 03 are referred as "Homo", "Hard $\beta'$", and "Soft $\beta'$". Details of these notations is given in Section 3.1.

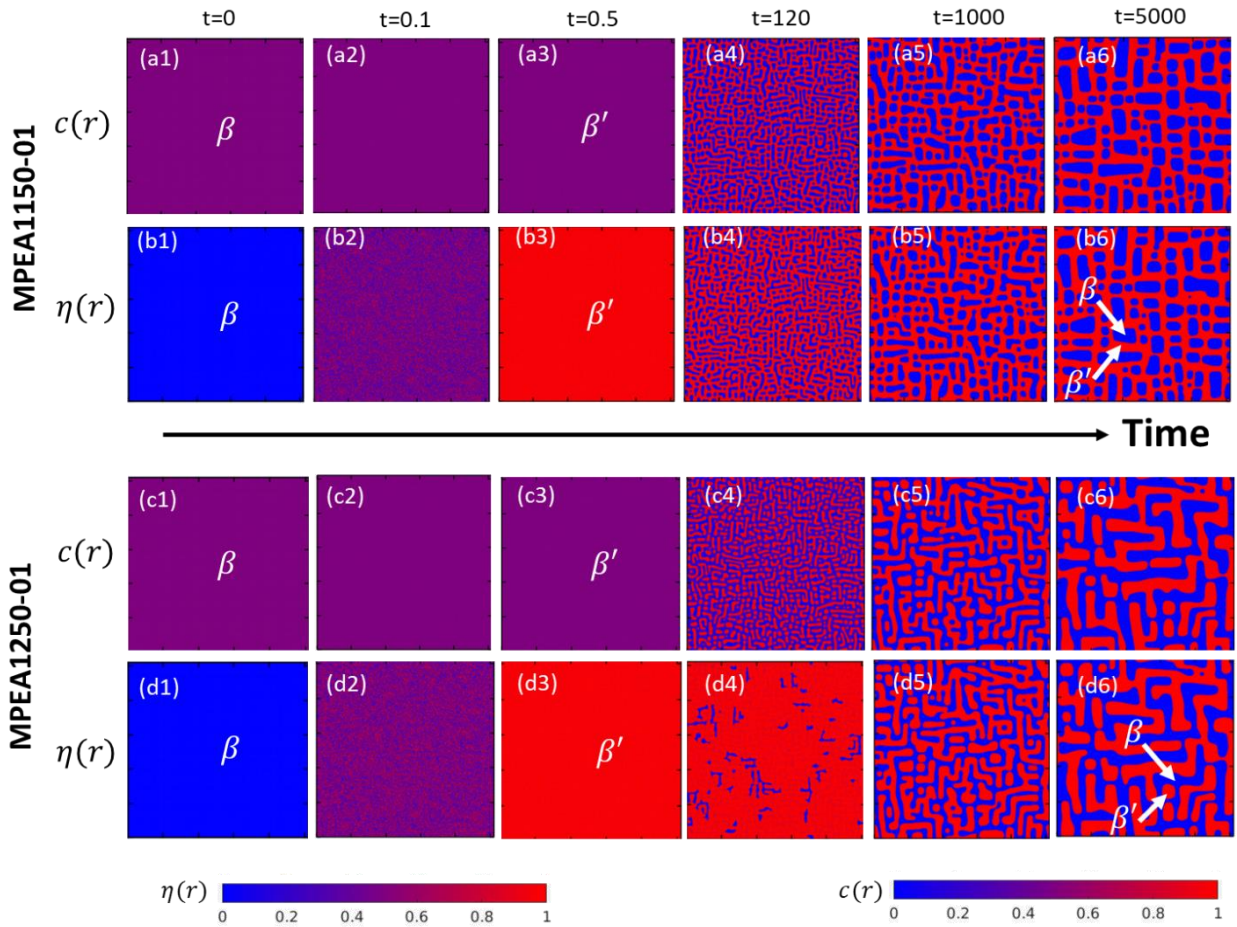

Figure 5. Microstructural evolutions in MPEA1150-01 and MPEA1250-01. Both alloys follow PTP (1) $\beta \rightarrow \beta' \rightarrow \beta'_1 + \beta'_2 \rightarrow \beta + \beta'_2$. These alloys have homogenous modulus, and the volume fraction ($f_V$) of the $\beta'$ is 50%.

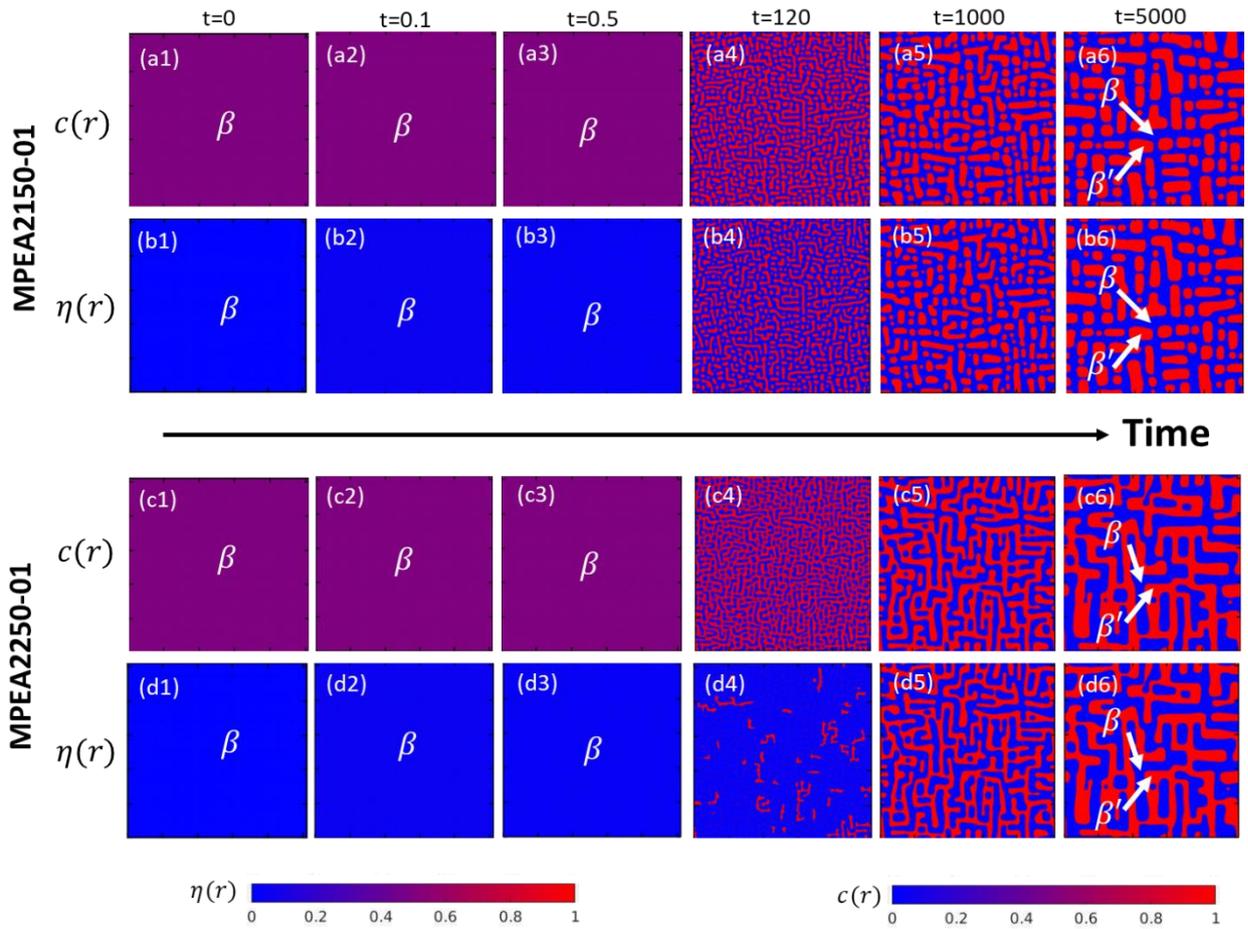

Figure 6. Microstructural evolutions of MPEA2150-01 and MPEA2250-01. Both alloys follow PTP (2): $\beta \rightarrow \beta_1 + \beta_2 \rightarrow \beta_1 + \beta'$. These alloys have homogenous modulus, and the volume fraction ($f_V$) of the $\beta'$ is 50%.

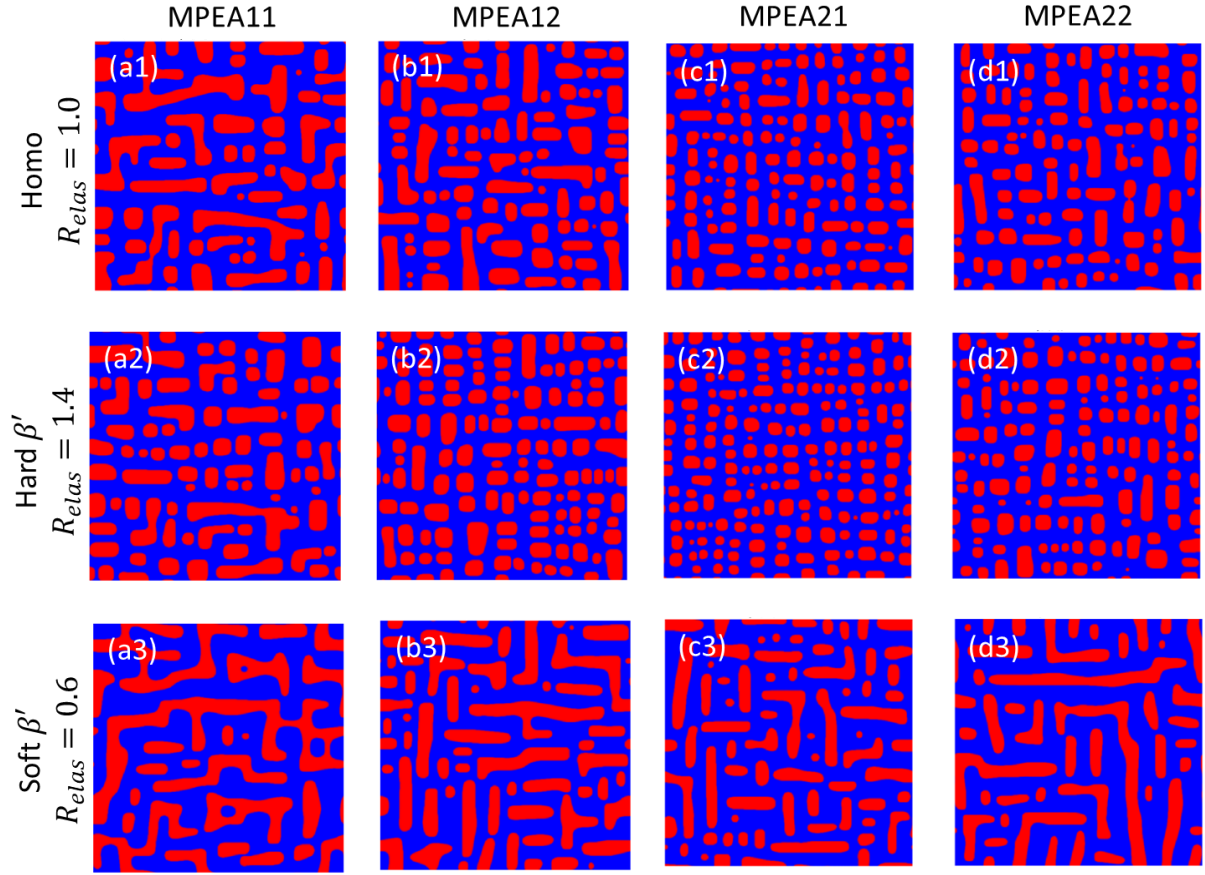

Figure 7. The final microstructures $c(\vec{r}, t)$ of the alloys with $f_V = 40\%$ volume fraction (MPEAXX40-ZZ). (a1) MPEA1140-01, (a2) MPEA1140-02, (a3) MPEA1140-03, (b1) MPEA1240-01, (b2) MPEA1240-02, (b3) MPEA1240-03, (c1) MPEA2140-01, (c2) MPEA2140-02, (c3) MPEA2140-03, (d1) MPEA2240-01, (d2) MPEA2240-02, (d3) MPEA2240-03. All the microstructures have $\beta$ matrix + $\beta'$ precipitate except for the bi-continuous microstructure of (a3) MPEA1140-03. Maximum discreteness is observed for (c2) MPEA2140-02.

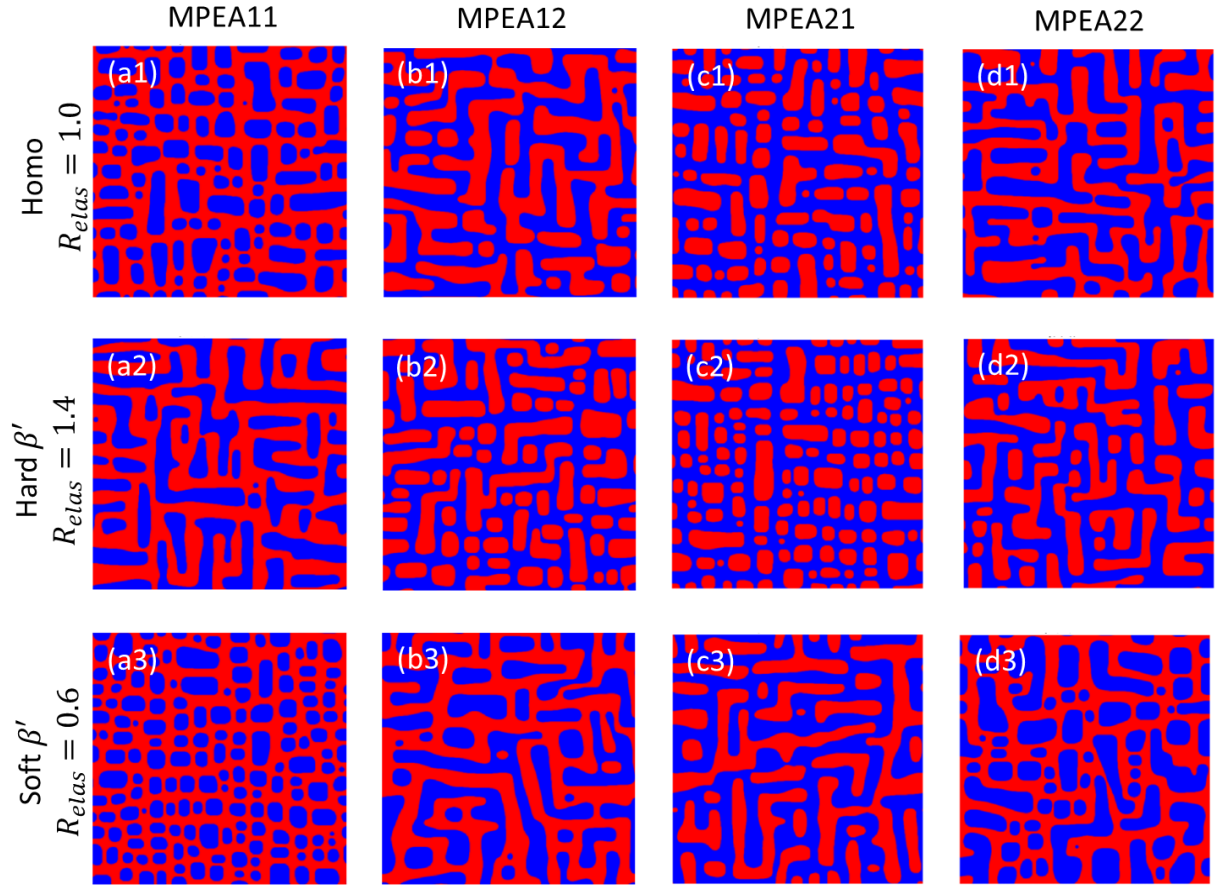

Figure 8. The final microstructures $c(\vec{r}, t)$ of the alloys with $f_V = 50\%$ volume fraction (MPEAXX50-ZZ). (a1) MPEA1150-01, (a2) MPEA1150-02, (a3) MPEA1150-03, (b1) MPEA1250-01, (b2) MPEA1250-02, (b3) MPEA1250-03, (c1) MPEA2150-01, (c2) MPEA2150-02, (c3) MPEA2150-03, (d1) MPEA2250-01, (d2) MPEA2250-02, (d3) MPEA2250-03. Both $\beta$ matrix and $\beta'$ matrix are observed depending on the free energy surface and modulus mismatch. Bi-continuous microstructure is observed for (d1) MPEA2250-01.

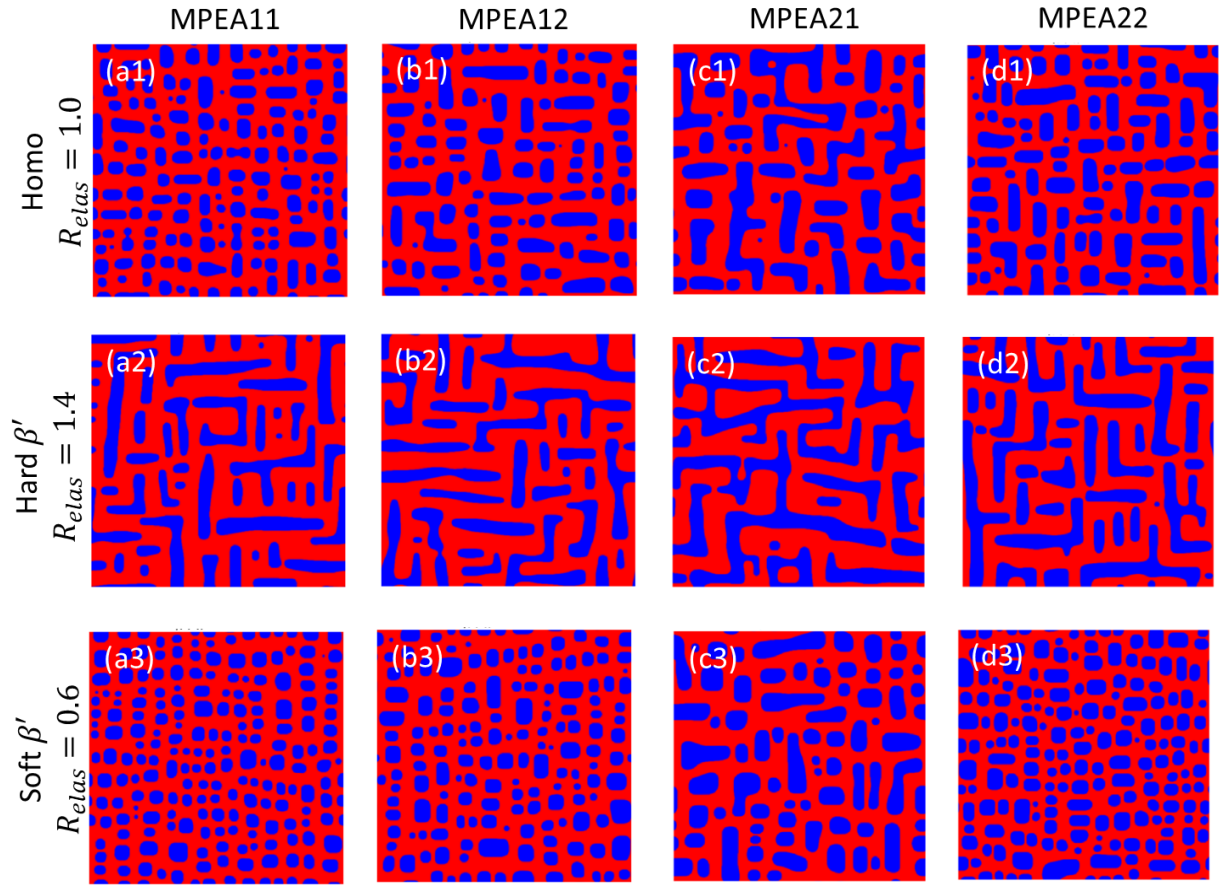

Figure 9. The final microstructures $c(\vec{r}, t)$ of the alloys with $f_V = 60\%$ volume fraction (MPEAXX60-ZZ). (a1) MPEA1160-01, (a2) MPEA1160-02, (a3) MPEA1160-03, (b1) MPEA1260-01, (b2) MPEA1260-02, (b3) MPEA1260-03, (c1) MPEA2160-01, (c2) MPEA2160-02, (c3) MPEA2160-03, (d1) MPEA2260-01, (d2) MPEA2260-02, (d3) MPEA2260-03. All the microstructures have $\beta'$ matrix + $\beta$ precipitate except for the bi-continuous microstructure of (c2) MPEA2160-02. Maximum discreteness is observed for (a3) MPEA1160-03.

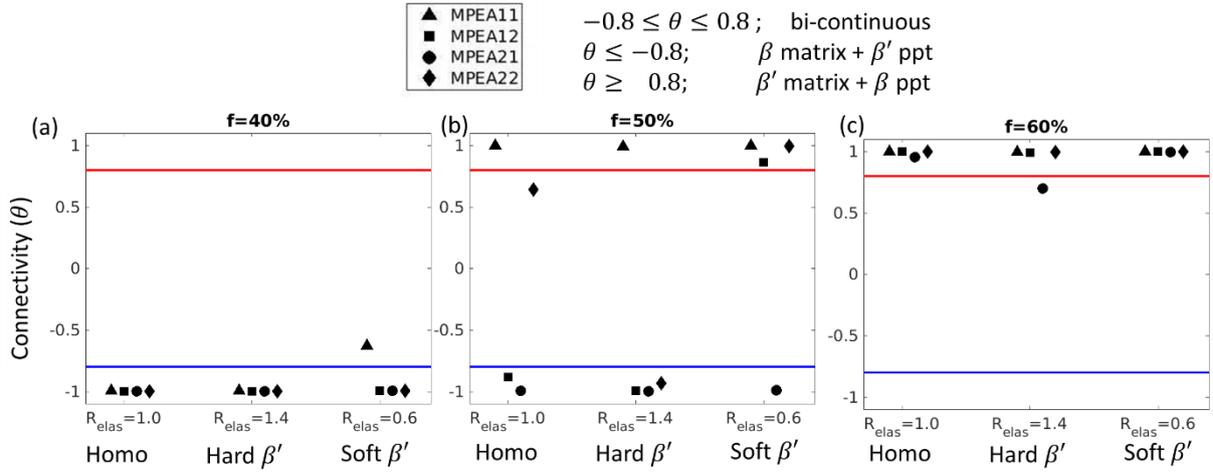

Figure 10. Connectivity $\theta$ of the microstructures for different volume fractions ($f_V$). (a) $f_V = 40\%$ (b) $f_V = 50\%$ (c) $f_V = 60\%$. The red and blue lines indicate the cutoff values of $\theta$ for solute-rich $\beta'$ matrix and solute-lean $\beta$ matrix respectively.

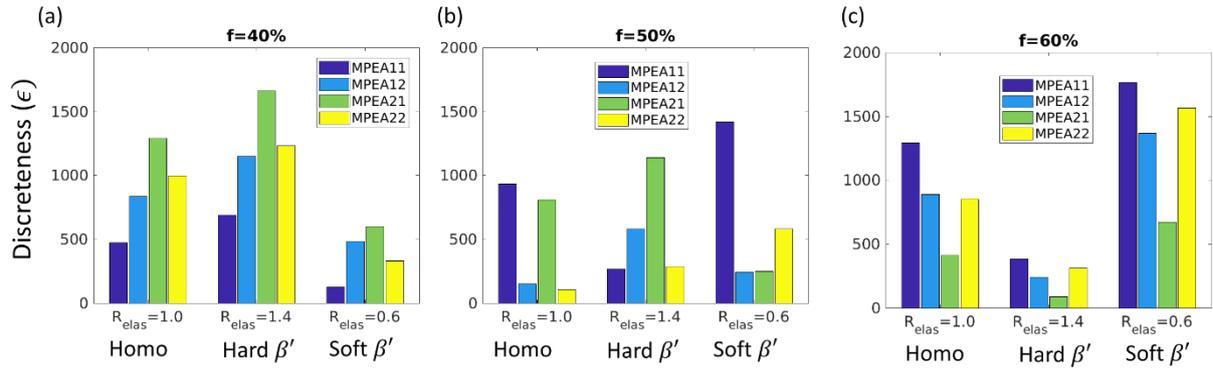

Figure 11. Discreteness $\epsilon$ of the microstructures for different volume fractions ($f_V$). (a) $f_V = 40\%$ (b) $f_V = 50\%$ (c) $f_V = 60\%$.

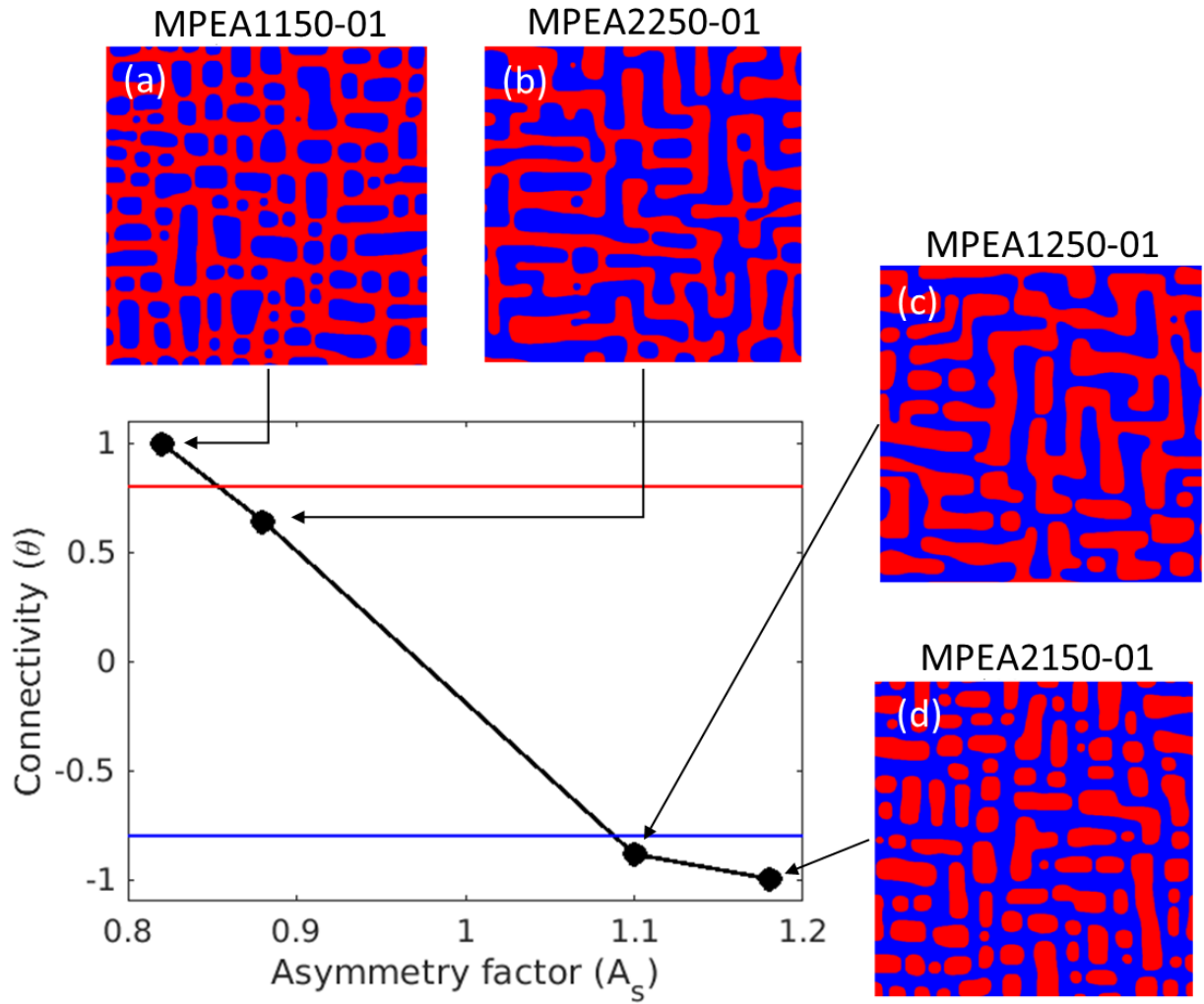

Figure 12. Effect of Asymmetry factor $A_s$ on Connectivity $\theta$. (a) MPEA1150-01 (b)MPEA2250-01 (c)MPEA120-01 (d)MPEA2150-01. In all these alloys, the volume fraction is $f_V = 50\%$ and the elasticity condition is $R_{elas} = 1.0$.

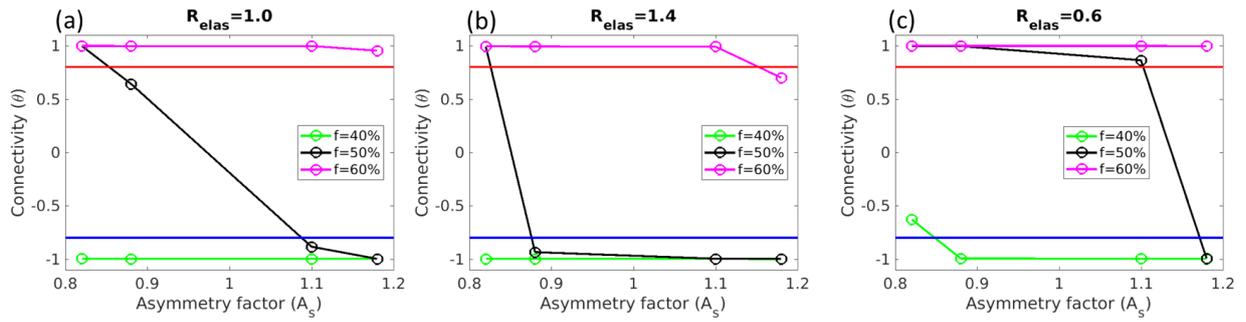

Figure 13. Connectivity $\theta$ as a function of $A_s$ for different elasticity conditions. (a)$R_{elas} = 1.0$; "Homo" (b)$R_{elas} = 1.4$; Hard $\beta'$ (c)$R_{elas} = 0.6$; Soft $\beta'$.

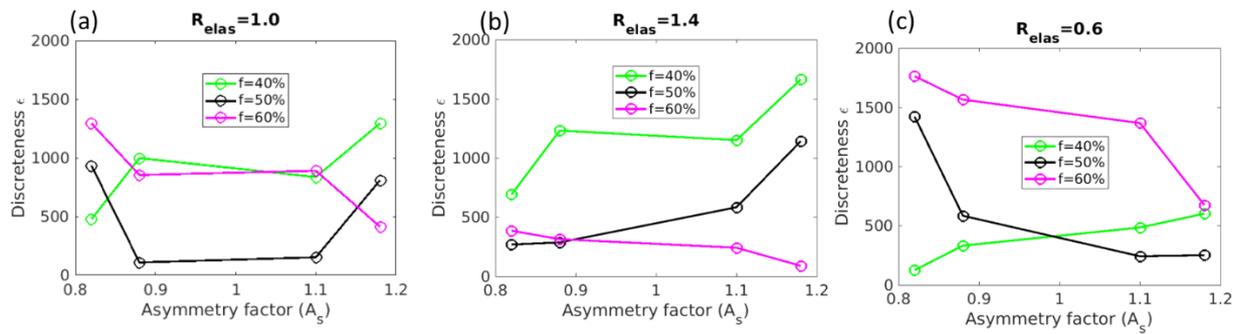

Figure 14. Discreteness $\epsilon$ as function of $A_s$ for different elasticity conditions. (a)$R_{elas} = 1.0$; "Homo" (b)$R_{elas} = 1.4$; Hard $\beta'$ (c)$R_{elas} = 0.6$; Soft $\beta'$.

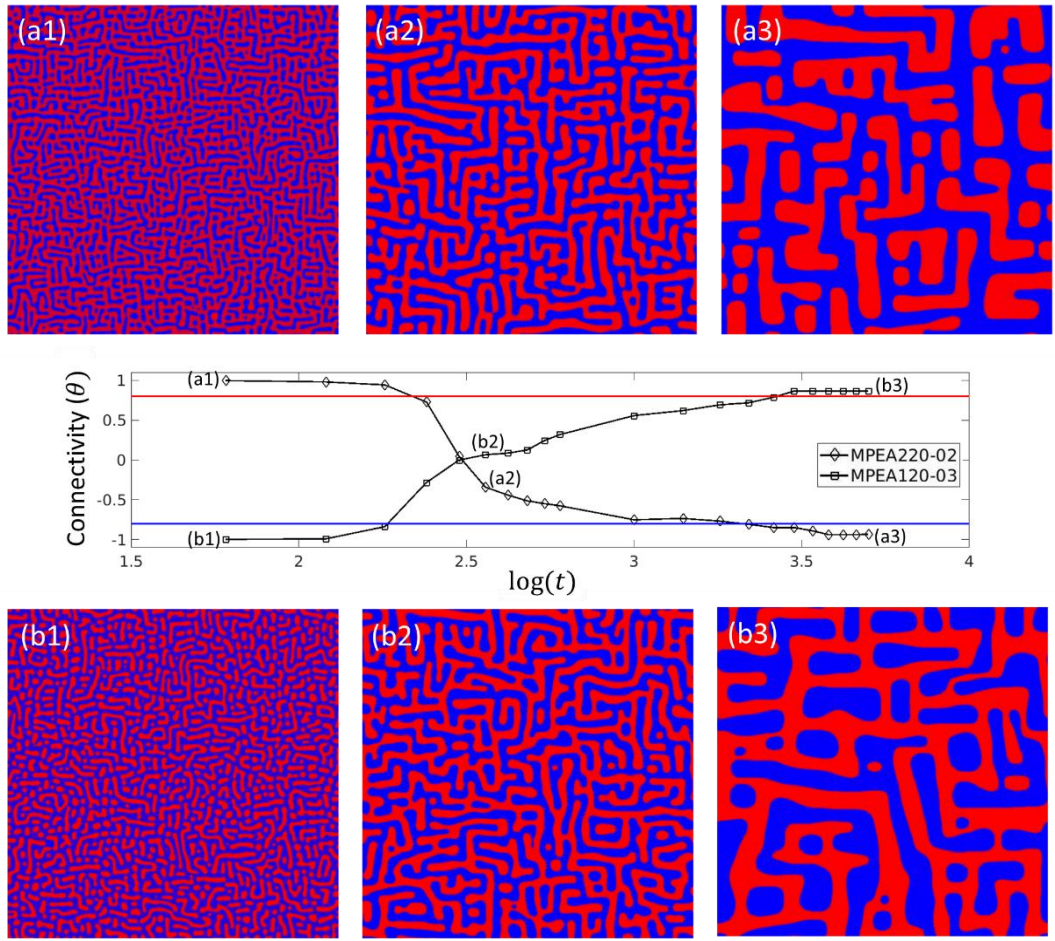

Figure 15. Connectivity ($\theta$) as a function of time ($t$) for the alloys MPEA2250-02 ($A_s = 0.88$) and MPEA1250-03 ($A_s = 1.1$). The microstructures (a1), (a2) and (a3) belong to MPEA2250-02. The microstructures (b1), (b2) and (b3) belong to MPEA1250-03. $\theta$ changes sign during the evolution for both the alloys. This phenomenon is termed as "phase inversion" [44].

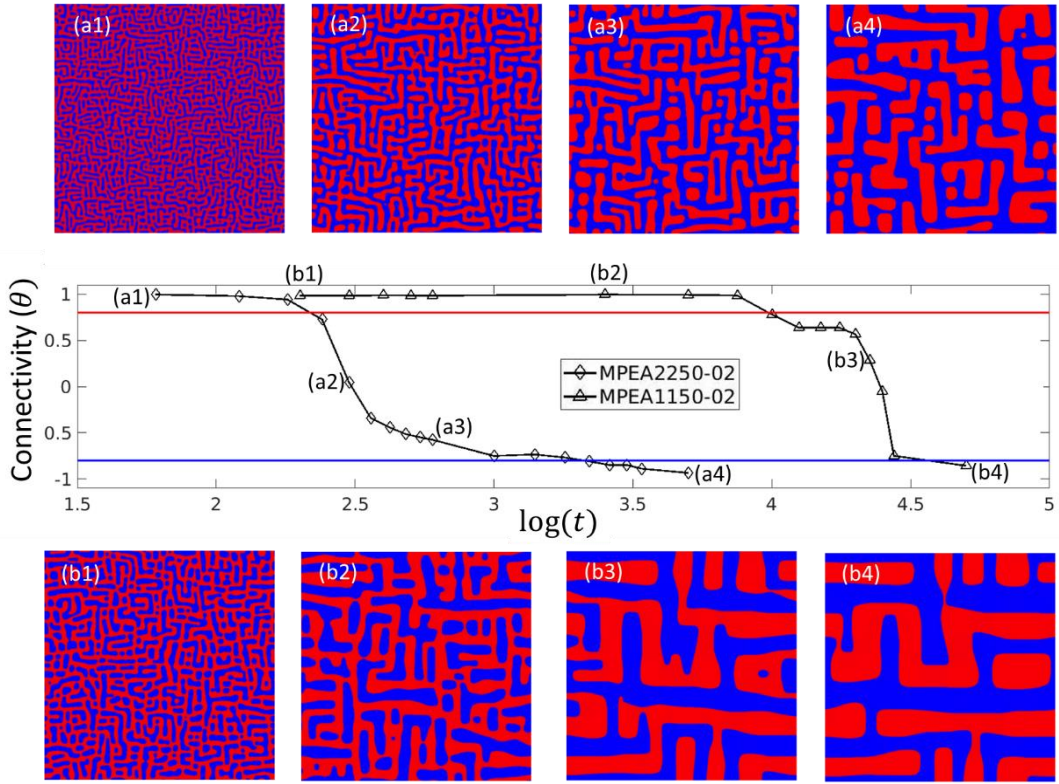

Figure 16. Effect of $A_s$ on time taken for phase inversion in equal volume fraction alloy, i.e., $f = 50\%$. (a) MPEA2250-02 ($A_s = 0.88$) (b) MPEA1150-02 ($A_s = 0.82$). MPEA1150-02 took longer time for phase inversion as it has higher $|A_s - 1|$ compared to MPEA2250-02.

**Supplementary Material**

Section A: Measured volume fraction at the end of the simulation ($t = 5000$)

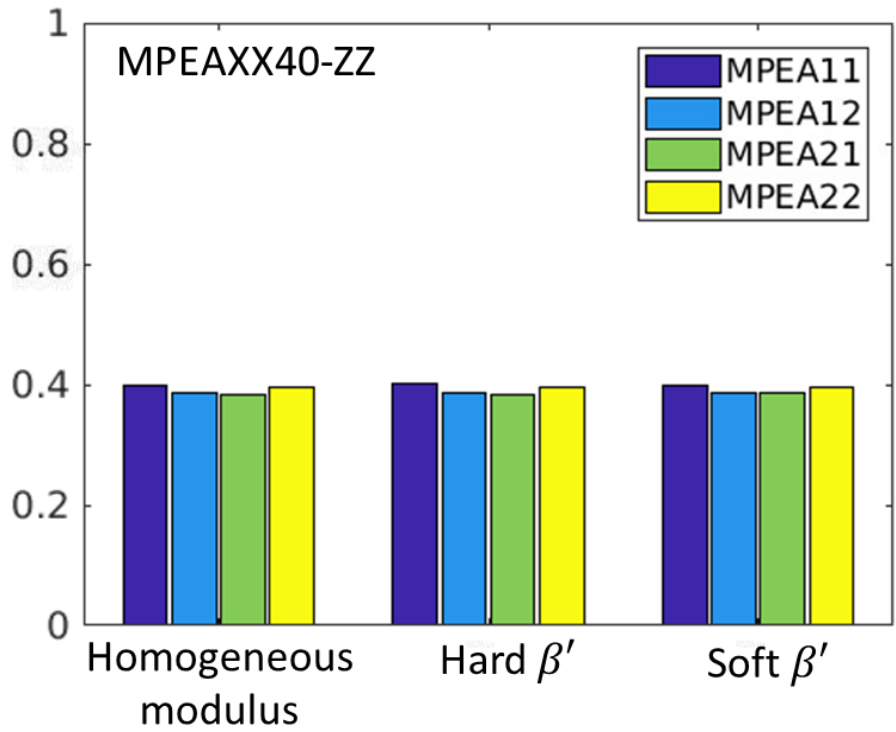

Figure S1. Measured volume fraction at the end of the simulation for the alloys with 40% equilibrium volume fraction.

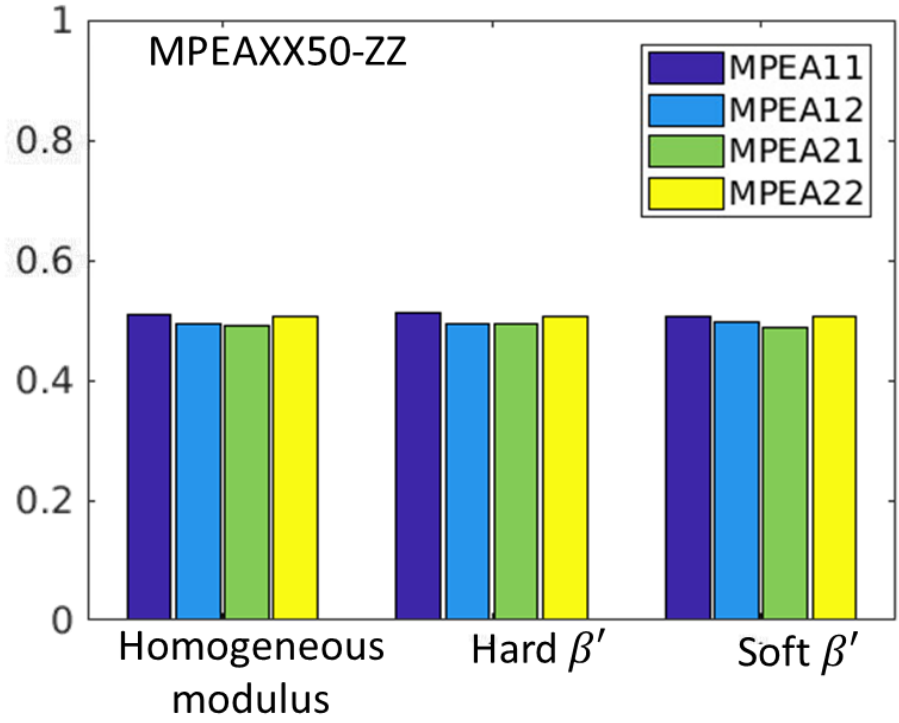

Figure S2. Measured volume fraction at the end of the simulation for the alloys with 50% equilibrium volume fraction.

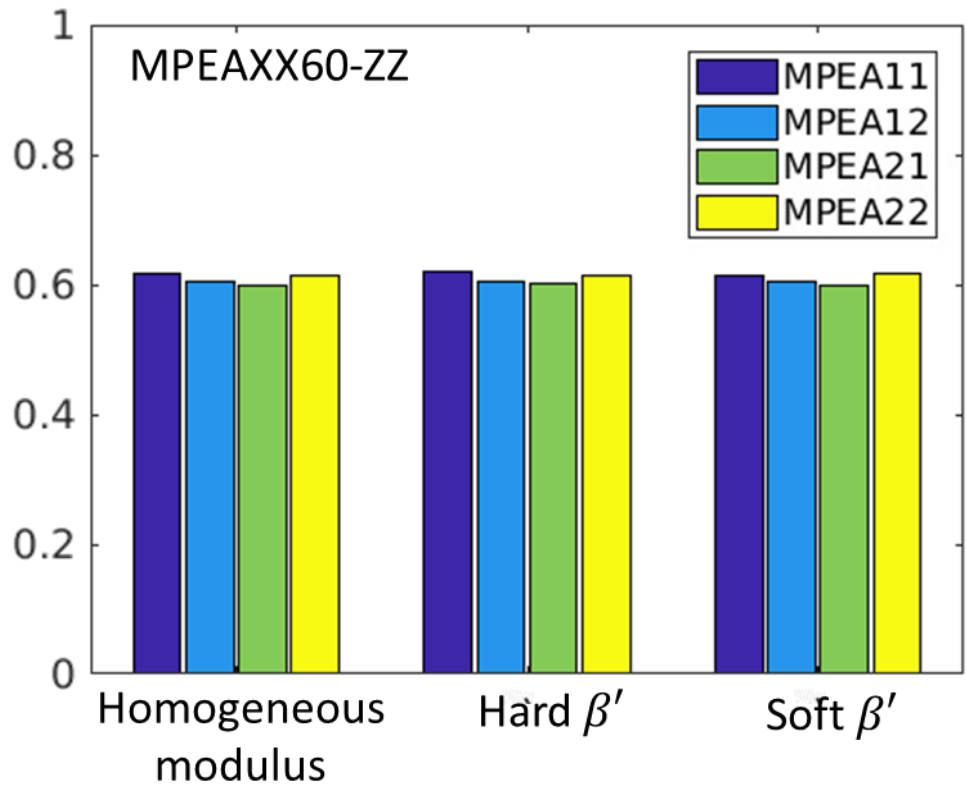

Figure S3. Measured volume fraction at the end of the simulation for the alloys with 60% equilibrium volume fraction.